\documentclass[fleqn,usenatbib]{mnras}
\usepackage{newtxtext,newtxmath}
\usepackage[T1]{fontenc}
\DeclareRobustCommand{\VAN}[3]{#2}
\let\VANthebibliography\thebibliography
\def\thebibliography{\DeclareRobustCommand{\VAN}[3]{##3}\VANthebibliography}

\usepackage{graphicx}	% Including figure files
\usepackage{amsmath}	% Advanced maths commands
\usepackage{pdflscape}

\usepackage{orcidlink}
\usepackage{balance}
\usepackage{dot2texi}
\usepackage{tikz}
\usetikzlibrary{shapes, arrows, positioning, backgrounds}
\usetikzlibrary{graphs, fit}
\usepackage{fix-cm}
\newcommand{\HI}{\ion{H}{i}}
\title[Deep uGMRT Observations for HERA Calibration]{Deep uGMRT observations for enhanced calibration of 21~cm arrays -- \\ I. First image and source catalogue}

\author[Elahi et al.]{Khandakar Md Asif Elahi$^{1}$\thanks{E-mail: asifelahi999@gmail.com (KMAE)}\,\orcidlink{0000-0003-1206-8689},
Samir Choudhuri$^{1}$\,\orcidlink{0000-0002-2338-935X},
Nirupam Roy$^{2}$\,\orcidlink{0000-0001-9829-7727},
Md Rashid$^{2,3}$\,\orcidlink{0009-0007-0120-5728}, 
Philip Bull$^{4,5}$\,\orcidlink{0000-0001-5668-3101}, \newauthor
and Dharam Vir Lal$^{6}$\\ %\,\orcidlink{0000-0000-0000-0000}
\\
$^{1}$ Centre for Strings, Gravitation and Cosmology, Department of Physics, Indian Institute of Technology Madras, Chennai 600036, India\\
$^{2}$ Department of Physics, Indian Institute of Science, Bangalore 560012, India \\
$^{3}$ Joint Astronomy Programme, Indian Institute of Science, Bangalore 560012, India \\
$^{4}$ Jodrell Bank Centre for Astrophysics, University of Manchester, Manchester, M13 9PL, United Kingdom\\
$^{5}$ Department of Physics and Astronomy, University of Western Cape, Cape Town 7535, South Africa \\
$^{6}$ National Centre for Radio Astrophysics, Tata Institute of Fundamental Research, Post Bag 3, Ganeshkhind, Pune - 411 007, India \\
}

\date{Accepted XXX. Received YYY; in original form ZZZ}

\pubyear{\the\year{}}

\begin{document}
\label{firstpage}
\pagerange{\pageref{firstpage}--\pageref{lastpage}}
\maketitle

% Abstract of the paper % 249 words now
\begin{abstract} 
    Radio-interferometric arrays require very precise calibration to detect the Epoch of Reionization 21-cm signal. A remarkably complete and accurate sky model is therefore needed in the patches of the sky used to perform the calibration. Instruments such as HERA, which use a redundant calibration strategy, also require a reference sky model to fix degenerate gain solutions. We have carried out a deep (20~hours) observation using the upgraded GMRT to make a high-fidelity sky model of one of the HERA calibration fields GLEAM~02H (J0200–3053). Here, we present the results from a $16.7\,\rm{MHz}$ bandwidth data centred at $147.4\,\rm{MHz}$. Using multiple GMRT pointings, we have made a $6.9^\circ\times6.9^\circ$ mosaic, which yields a median rms of $3.9^{+3.7}_{-1.4}$~mJy/beam that reduces to $\sim2$~mJy/beam at source-free regions. In the overlapping patch, this rms is deeper than the GLEAM catalogue, which is used for HERA calibration. We produce a catalogue of $640$ sources ($26\%$ extended) in the flux range $0.01-19.08$~Jy. The catalogue has a sub-arcsec positional accuracy, and the estimated fluxes are consistent with existing catalogues. The differential source counts are found to be deeper than GLEAM and consistent with LoTSS. Preliminary simulations of the sky models from GLEAM and our catalogue show $\sim 10-25\%$ difference in the visibility amplitude, with relatively small phase difference ($\approx 2^\circ$). Future work is planned for larger survey areas and wider bandwidth to reduce the rms and measure the in-band source spectral indices, which are expected to enhance the fidelity of the HERA calibration model.     
\end{abstract}

\begin{keywords}
instrumentation: interferometers -- methods: data analysis -- cosmology: observations, dark ages, reionization, first stars -- radio continuum: general
\end{keywords}

\section{Introduction}

Epoch of Reionization (EoR) is a crucial phase in the history of our Universe when the X-ray and UV emission from the first stars and galaxies ionized the neutral Hydrogen (\HI{}) of the intergalactic medium. The 21-cm signal emitted by the \HI{} during this period can reveal the timing and duration of reionization, constrain the properties of ionizing sources such as the first stars, galaxies, and quasars, and provide insights into the morphology of reionization, including the typical sizes and distribution of ionized bubbles \citep{Madau1997, Barkana2001,  Bharadwaj2005, Furlanetto2006, Pritchard2012, mellema13, Ghara2024IGM, Giri2024, Mishra2025}. 

Measuring the EoR 21-cm signal is one of the most ambitious goals in modern observational cosmology. The Giant Metrewave Radio Telescope (GMRT; \citealt{Swarup1991, Gupta2017}), the Murchison Widefield Array (MWA; \citealt{Tingay2013}), the LOw Frequency ARray (LOFAR; \citealt{Zaroubi2012, vanHarlem2013}), the Hydrogen Epoch of Reionization Array (HERA; \citealt{DeBoer2017}), and the Square Kilometre Array (SKA-Low; \citealt{mellema13, Koopmans2015}) are the radio interferometers that are trying to measure the EoR 21-cm power spectrum. The major challenges in these experiments are foreground emission from our Galaxy and other galaxies, which are many orders of magnitude brighter than the expected signal \citep{ali, bernardi09, ghosh3}, and the systematics that arise due to the interaction of the foregrounds with the instrumental response \citep{adatta10, ghosh1, Morales2012, vedantham12, trott2012, parsons12, liu14a, liu14b, pober16, Murray_2018, Pal2020}. Extracting the weak cosmological 21-cm signal from the observed data requires precise calibration of the interferometric array \citep{Barry16, jais2020, jais22, Gayen_2025}, accurate foreground modelling \citep{ghosh3}, and robust statistical techniques \citep{mertens18, Kennedy2023}. Despite extensive efforts, the EoR 21-cm power spectrum has not been detected yet, and currently, we only have upper limits using different radio interferometers \citep{Paciga2013, Kolopanis2019, Mertens2020, Trott2020, Pal2020, patwa2021, Abdurashidova2022, Kolopanis2023, Acharya2024, Chatterjee2024, Mertens2025, Elahi2025}. 

HERA is currently the most sensitive telescope in the world dedicated to the statistical detection of the 21-cm signal. It recently published the tightest $2\sigma$ upper limit on the 21-cm power spectrum $\Delta^2(k) < (21.4)^2\,{\rm mK}^2$ at $k = 0.34\, h\,{\rm Mpc}^{-1}$ at $z = 7.9$ \citep{HERA2023}. The full-array configuration of HERA is expected to detect the power spectrum with high significance \citep{DeBoer2017}. A recent expansion of HERA’s frequency range to 50–250~MHz (see \citealt{Fagnoni2021Herap2}) has also enhanced its ability to probe the Cosmic Dawn  - the epoch when the very first stars and galaxies were formed. The expansion includes the frequency band around 78~MHz, at which an anomalous absorption profile, associated with the sky-averaged (`global') 21-cm signal, was measured by EDGES \citep{Bowman2018}. However, the absorption profile was not found in an independent experiment, SARAS \citep{Singh2022a}, and it still remains an open problem.

Detecting the 21-cm signal requires unprecedented precision in calibration of the instrument. A `redundant' array, such as HERA, has a distinctive design that makes it a unique instrument for 21-cm experiments. It has a highly regular hexagonal close-packed array layout that provides many baselines with almost-identical lengths and orientations. This allows for a redundant calibration strategy in which the nearly identical baselines measure the same sky signal, and it allows one to find robust relative gain solutions \citep{Wieringa1992, Liu2010}. However, this technique leaves four degenerate parameters (per frequency, time, and polarizations) that must be resolved using an external sky model, known as the absolute calibration process \citep{Dillon2018:redcal, Dillon2020, Kern2020}.

The precision of absolute calibration is fundamentally tied to the quality of the sky models and the instrument. The faint point sources, which are not present in the model due to the limitation in the sensitivity, can introduce spurious spectral structure in the gain solutions, causing foreground leakage into the clean EoR window. We ideally need to achieve a calibration accuracy of $\sim10^{-5}$ \citep{Barry16}, which is challenging due to errors in source flux densities, spectral indices, and missing sources in existing catalogues. Even with a perfect redundant calibration, the incompleteness in the sky model introduces foreground leakage in the EoR window that can limit detection of the 21-cm signal \citep{Dillon2018:redcal}. The accuracy of the reference sky model, therefore, plays a crucial role in achieving high-precision calibration. \cite{Byrne2019} showed that a better sky model, in terms of accuracy and completeness, can minimize errors from the sky model incompleteness. As a direct example, \cite{Ewall-Wice2017} used simulations to show that incorporating the TIFR-GMRT Sky Survey (TGSS; \citealt{Intema17}) source catalogue can significantly improve HERA’s calibration. They further determined that modelling sources down to 0.1~mJy using long baselines would eliminate systematic calibration errors that currently limit EoR detection. However, TGSS and similar surveys do not have the required sensitivity and accurate estimates of the spectral indices for many sources, which again complicates absolute calibration.  

Recent HERA results \citep{Abdurashidova2022, HERA2023} have used the GLEAM (GaLactic and Extragalactic All-sky MWA Survey) catalogue \citep{Wayth2015, HW17} to construct the reference sky model for absolute calibration. The GLEAM catalogue covers $24,831$ deg$^2$ on the sky over declinations south of $+30^{\circ}$ and it contains $307~455$ radio sources whose flux densities are measured across the bandwidth $72-231$ MHz. Considering the region $-72^{\circ} \leq {\rm Dec} < +18.5^{\circ}$, which has an overlap with sky that is observed by HERA, GLEAM has achieved a root-mean-square (rms) sensitivity of $10 \pm 5$~mJy/beam from a wideband image produced at 200~MHz. For the same frequency and the sky coverage, the catalogue is estimated to be $\sim90\%$ complete at $100$~mJy, and $50\%$ complete at $\sim50$~mJy, and has a reliability of $99.97\%$ above the detection threshold of $5\sigma$ ($\sim50$~mJy). Recently, GLEAM--X \citep{HW22}, which uses the `extended' Phase~II configuration of MWA, has significantly improved sensitivity over the previous GLEAM catalogue. The second data release of the GLEAM--X survey \citep{Ross2024} covers an area of $12~892\,\mathrm{deg}^2$ in the range $20^h40^m \leq \mathrm{RA} \leq 6^h40^m, -90^{\circ}  \leq \rm{Dec} \leq +30^{\circ}$. The catalogue comprises $624~866$ components, which include $562~302$ spectrally fitted components. From a wideband (170–231~MHz) mosaic, GLEAM--X has achieved a median value of the rms $1.5^{+1.5}_{-0.5}$~mJy/beam. Further, the catalogue is estimated to be $\sim90\%$ complete at $10.2$~mJy, and $50\%$ complete at $\sim5.8$~mJy, and has a reliability of 98.7\% at a $5\sigma$ level.  

The upgraded GMRT (uGMRT) offers several advantages that make it an ideal instrument for improving the GLEAM-based reference sky model used in HERA’s absolute calibration. GMRT has much longer baselines than MWA, which yields significantly improved angular resolution, and it also helps to beat down confusion noise. The recent upgradation enables GMRT to have a nearly continuous frequency coverage over the range $120 - 1500$~MHz, with large instantaneous bandwidths, which provides a better handle to measure the spectral indices of the sources detected with high resolution. Further, the uGMRT has an improved receiver with low system noise and better dynamic range than the legacy GMRT, and this makes it highly sensitive for continuum studies over a large frequency range until SKA  becomes operational \citep{Gupta2017}. With its superior angular resolution and sensitivity, the uGMRT has the potential for deeper imaging of HERA calibration fields, thereby yielding more complete source models. These improved models can be integrated into HERA’s absolute calibration pipeline to mitigate systematic errors and enhance 21-cm power spectrum measurements. Additionally, the smaller field of view of uGMRT is expected to reduce the influence of the very bright extended sources like Pictor A and Fornax A, which appear in the MWA and HERA sidelobes. The complementary measurements with uGMRT offer a different Radio Frequency Interference (RFI) environment, receiver system characteristics, and antenna shapes (dish/dipoles), which present an opportunity to cross-check the source properties between the MWA and uGMRT detections to ensure that we have a good handle on systematic error contributions to the sky model. In addition to improving the HERA absolute calibration, uGMRT's complementary baseline layout and deep observations also promise to yield tight upper limits with uGMRT itself at the (smaller) length scales that are inaccessible to HERA.

To demonstrate the idea that we can actually improve the sky model for the HERA calibration, we have carried out deep (20~hours) uGMRT observations targeting one of the chosen calibration fields, GLEAM~02H (J0200–3053). In this first paper of this series, we focus on establishing the framework for building the sky model and demonstrate the feasibility of using actual uGMRT observations to improve HERA calibration. Future works are planned for increasing the survey areas and bandwidth to systematically build up an enhanced sky model, and thereby reduce systematics in 21-cm power spectrum measurements.  

The rest of the paper is organized as follows: Section~\ref{sec:obs&proc} describes the uGMRT observations and the data processing.  Section~\ref{sec:images} shows the details of the images and the background rms noise that we obtained. Section~\ref{sec:catalog} presents the source catalogue and its cross-match with existing catalogues. We compute the Euclidean normalized differential source count in Section~\ref{sec:sourcecount} to demonstrate the depth of the catalogue. In Section~\ref{sec:herasim}, we use the uGMRT and GLEAM catalogues to simulate sky models for HERA and compare these. Our findings are summarized in Section~\ref{sec:conclusions}.

\section{uGMRT Observation and Data Processing}
\label{sec:obs&proc}

\subsection{Observation}
\label{sec:ugmrt_observation}

% \subsection{Choice of HERA Calibration Fields}
\begin{figure*}
    \centering
    \includegraphics[width=\textwidth]{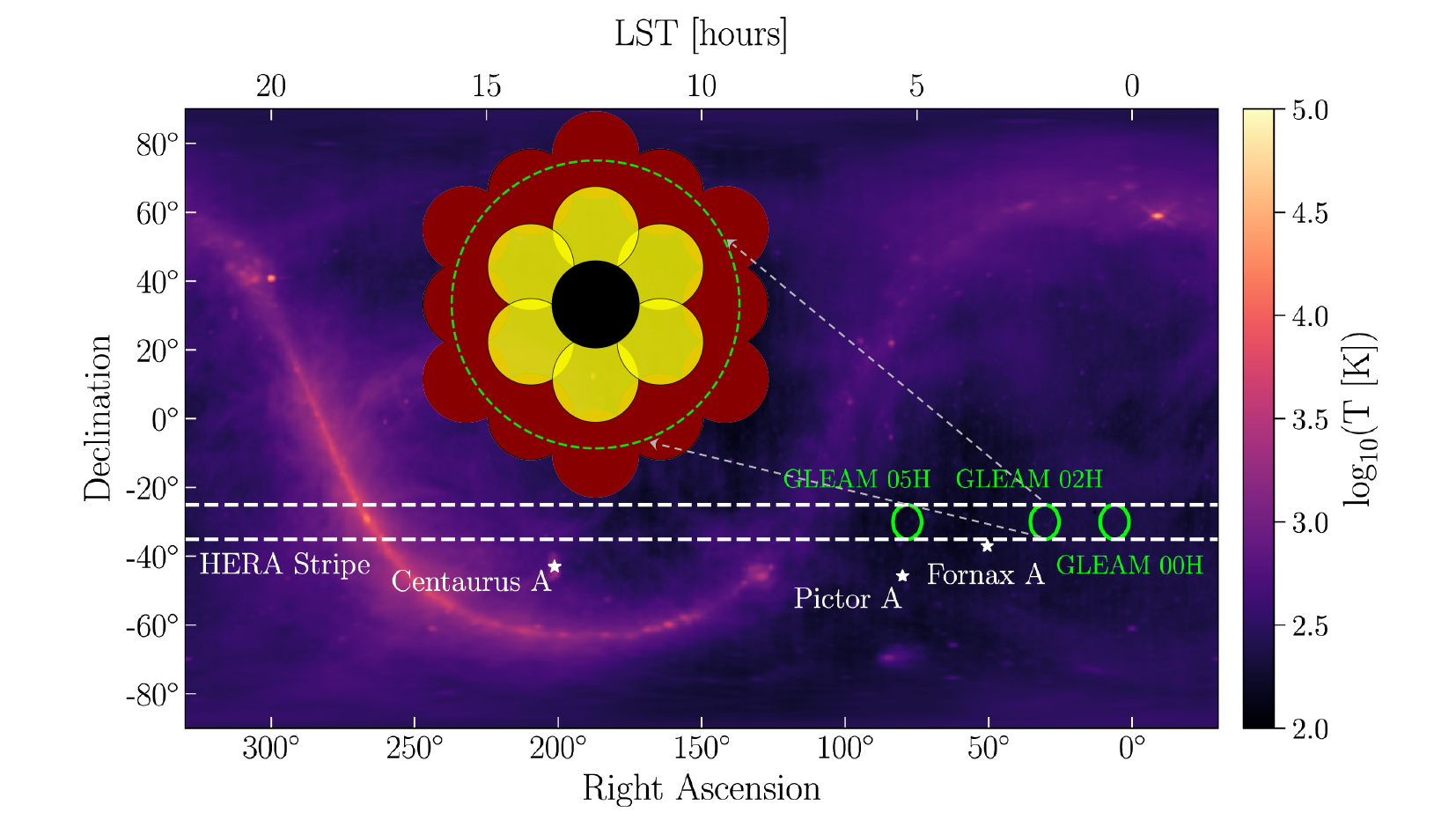}
    \caption{The radio sky at 150 MHz from the Global Sky Model \citep{DeOliveiraCosta2008}. The dashed lines show the HERA observation stripe centred at HERA’s latitude of $-30.7^{\circ}$ with a width of $10^{\circ}$, which is the FWHM of the primary beam at 150~MHz. The green solid circles show the three ideal calibration fields currently used for absolute calibration. The inset depicts our strategy to observe the GLEAM~02H field with the uGMRT. Adapted from \citet{Kern2020}.}
    \label{fig:herastrip}
\end{figure*} % GLEAM~02H (RA = 02$^{h}$ 00$^{m}$ 12.00$^{s}$ and Dec = -30$^{\circ}$ 53$^{\prime}$ 23.99$^{\prime\prime}$)

\begin{figure}
\centering
    \includegraphics[width=\columnwidth]{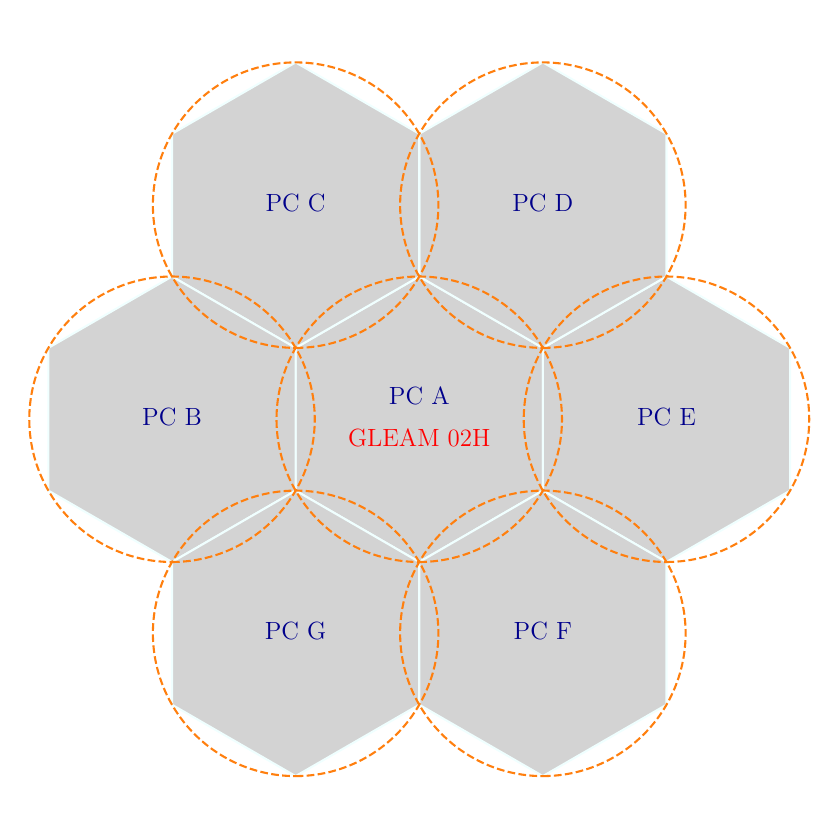}
    \caption{The strategy to observe the field GLEAM~02H with uGMRT.  The dashed circles represent the Full Width at Half Maxima (FWHM = $1.82^{\circ}$) at 250~MHz, the highest available frequency of Band 2 ($120-250$~MHz). The positions (RA, Dec) of all seven PCs are in Table~\ref{table:ra_dec}.}
    \label{fig:observation_strategy}
\end{figure}

We have carried out GMRT observations to make a highly complete and accurate sky model that can be used to improve the calibration of a complementary telescope such as HERA. Recently, GMRT has been upgraded to improve its bandwidth to achieve a near-seamless coverage of the frequency range $120-1500$~MHz \citep{Gupta2017}. This makes the upgraded GMRT (hereafter uGMRT) suitable for making high-resolution images along with the leverage of accurate measurement of spectral indices of the sources. We have used the Band~2 ($120-250$~MHz) frequency range, which overlaps with HERA observing frequency. Our observations were recorded in both the GMRT Software Backend (GSB) and GMRT Wideband Backend (GWB) backends. The present work makes use of only the GSB data to establish the full framework for building the sky model and demonstrate the feasibility of using actual uGMRT observations to improve HERA calibration. % A detailed analysis of the GWB data is planned for a follow-up study.

Figure~\ref{fig:herastrip} (adapted from Figure 3 of \citealt{Kern2020}) shows the radio sky at 150 MHz from the Global Sky Model \citep{DeOliveiraCosta2008}. Here, the dashed lines show the HERA observation stripe centred at HERA’s latitude of $-30.7^{\circ}$ with a width of $10^{\circ}$, which is the full width at half maximum (FWHM) of the primary beam at 150~MHz. To perform an absolute calibration of HERA, it is necessary to have an isolated bright source near ${\rm Dec} = -30.7^{\circ}$. \cite{Kern2020} used the GLEAM catalogue to identify three such sources at the local sidereal times (LSTs) of  $0^h$ (J0024–2929), $2^h$ (J0200–3053), and $5^h$ (J0455–3006), which are referred to as GLEAM~00H, 02H, and 05H, respectively. These three fields, which are shown with green solid circles in Figure~\ref{fig:herastrip}, have at least one bright ($>15$~Jy at 150~MHz) point source that passes close to the beam centre. We note that \citet{jacobs2016hera} also identified GLEAM~02H in the TGSS-ADR catalogue \citep{Intema17} as a suitable source for the HERA calibration. Here we observe the GLEAM~02H with the uGMRT with higher resolution and sensitivity.

The inset of Figure~\ref{fig:herastrip} shows our strategy to observe the GLEAM~02H field with the uGMRT. Here, each of the filled circles produces uniform coverage in a nearly $3^\circ$ circular region. Six filled circles (yellow), around the central one (black), give a coverage of around $6^\circ$, whereas 12 additional filled circles (red) yield uniform sensitivity over the full $10^\circ$ FoV of HERA, which is shown by the dashed circle. We ideally want to have deep observations along with uniform coverage over the entire $10^\circ$ region around GLEAM~02H at the entire $120-250$~MHz frequency range of uGMRT Band 2. However, the GMRT primary beam has an FWHM of approximately $1.82^{\circ}$ at the upper end of this band ($250$~MHz), and therefore each of the filled circles itself requires multiple GMRT pointings for a uniform coverage. Focusing solely on the innermost filled circle (dark black), we have carried out a pilot survey through the GMRT proposal $43\_119$\footnote{\url{https://naps.ncra.tifr.res.in/goa/}}. 
As shown in Figure~\ref{fig:observation_strategy}, we have used 7 pointing centers (PCs) to uniformly cover this $3^\circ$ area at 250~MHz. Since the FWHM increases at lower frequencies, this configuration ensures full coverage of the target region across the entire frequency range of Band~2.  We label the pointing centres (PCs) as PC~A, PC~B, etc., where the central pointing (PC~A) coincides with GLEAM~02H. The right ascension (RA) and declination (Dec) of the PCs are listed in Table~\ref{table:ra_dec}. We note that we have recently proposed (ID: $49\_019$) a larger sky survey to cover the full $10^\circ$ sky around GLEAM~02H, and another 40~hours of observation time has been awarded for this purpose.

We have carried out $\sim$1.8 hours of on-source integration per PC, which results in a total on-source time of $\sim$12.4 hours. Additionally, we have observed the primary (flux and bandpass) calibrator 3C~48 at the start, middle, and end of each observing session, for a total of $\sim$2.3 hours, and the phase calibrator PMN~0116--2052\footnote{The calibrator is identifiable with its J2000 coordinates 0116--208 in the VLA calibrator list \url{https://science.nrao.edu/facilities/vla/observing/callist}.} intermittently for a total of $\sim$2.2 hours. Including overheads, this utilizes the full 20 hours allocated to us under GMRT proposal ID $43\_119$. The entire observation has been conducted over three nights, from 17 to 20 November 2022. A summary of the pointing centres and observational parameters is provided in Tables~\ref{table:ra_dec} and \ref{tab:obs_summary}.

To ensure good $uv$-plane sampling, we have observed different PCs at staggered times across the three nights. For instance, PCs~A -- G have been observed sequentially on the first night. On the next night, the sequence has been altered to observe PCs~G and~F first, followed by A -- E. On the third night, PCs~C and~D are observed first, followed by E -- G, and finally B and~A. Each observing sequence has been repeated twice per night. The resulting $uv$-coverages and related discussions are provided in Appendix~\ref{sec:uvtracks}.

\begin{table*}
    \centering
    \caption{RA and Dec values for different fields of the observation, along with observation time, flagging percentage, theoretical rms, off-source rms, and the major axis (Maj), minor axis (Min), and position angle (PA) of the FWHM of the PSF. Considering PC~A, the data taken on 17 November are completely flagged. The source 3C~48 has been used for flux and bandpass calibration, and PMN~0116--2052 is used as a phase calibrator.} 
    \label{table:ra_dec}
    \renewcommand{\arraystretch}{1.4} 
    \setlength{\tabcolsep}{5pt} 
    \begin{tabular}{cccccccc}
    \hline
    Fields & RA & Dec & Obs. Time & Flagging  & Theoretical rms & rms (median) & Maj $\times$ Min, PA \\
    & (h m s) & (d m s) & (hours) & (\%) & (mJy) & (mJy/beam) & (arcsec $\times$ arcsec, deg) \\
    \hline
    PC~A      & 02$^{h}$ 00$^{m}$ 12.00$^{s}$ & --30$^{\circ}$ 53$^{\prime}$ 23.99$^{\prime\prime}$ & 1.77 & 54 & 0.44 & 2.951 & $29.7^{\prime\prime} \times 14.8^{\prime\prime}, 23^\circ$ \\
    PC~B      & 02$^{h}$ 06$^{m}$ 31.11$^{s}$ & --30$^{\circ}$ 53$^{\prime}$ 23.99$^{\prime\prime}$ & 1.77 & 30 & 0.35 & 2.602 & $39.3^{\prime\prime} \times 14.6^{\prime\prime}, 23^\circ$) \\
    PC~C      & 02$^{h}$ 03$^{m}$ 21.56$^{s}$ & --29$^{\circ}$ 31$^{\prime}$ 19.19$^{\prime\prime}$ & 1.77 & 30 & 0.35 & 2.324 & $32.7^{\prime\prime} \times 14.9^{\prime\prime}, 12^\circ$ \\
    PC~D      & 01$^{h}$ 57$^{m}$ 02.44$^{s}$ & --29$^{\circ}$ 31$^{\prime}$ 19.19$^{\prime\prime}$ & 1.77 & 30 & 0.35 & 2.533 & $30.0^{\prime\prime} \times 15.2^{\prime\prime}, 13^\circ$ \\
    PC~E      & 01$^{h}$ 53$^{m}$ 52.89$^{s}$ & --30$^{\circ}$ 53$^{\prime}$ 23.98$^{\prime\prime}$ & 1.77 & 29 & 0.35 & 2.514 & $31.9^{\prime\prime} \times 14.8^{\prime\prime}, 22^\circ$ \\
    PC~F      & 01$^{h}$ 57$^{m}$ 02.44$^{s}$ & --32$^{\circ}$ 15$^{\prime}$ 28.79$^{\prime\prime}$ & 1.77 & 29 & 0.35 & 2.978 & $32.4^{\prime\prime} \times 16.7^{\prime\prime}, 25^\circ$ \\
    PC~G      & 02$^{h}$ 03$^{m}$ 21.56$^{s}$ & --32$^{\circ}$ 15$^{\prime}$ 28.79$^{\prime\prime}$ & 1.77 & 30 & 0.35 & 2.520 & $34.9^{\prime\prime} \times 15.5^{\prime\prime}, 15^\circ$ \\
    3C 48      & 01$^{h}$ 37$^{m}$ 41.67$^{s}$ & +33$^{\circ}$ 05$^{\prime}$ 31.72$^{\prime\prime}$ & 2.32 & -- & -- & -- & -- \\
    PMN~0116--2052 & 01$^{h}$ 16$^{m}$ 51.51$^{s}$ & --20$^{\circ}$ 52$^{\prime}$ 06.22$^{\prime\prime}$ & 2.22 & -- & -- & -- & -- \\
    \hline
    \end{tabular}
\end{table*}
% 0116-208   J2000  C 01h16m51.4043s   -20d52'06.813"
 % 0116-208 (PKS J0116-2052)

% \subsection{observation time}
\begin{table}
    \centering
    \renewcommand{\arraystretch}{1.2} 
    \setlength{\tabcolsep}{20pt} 
    \caption{Observation summary for GSB data.}
    \label{tab:obs_summary}
    \begin{tabular}{lc}
        \hline
        Parameters & Values \\
        \hline
        Working antennas & 28--29 \\
        Central Frequency & 147.4~MHz \\
        Bandwidth & 16.7~MHz \\
        Frequency resolution & 65~kHz \\
        Integration time & 2~sec \\
        Total Observation time & 20~hours \\
        Overhead time & 3~hours \\
        Calibrator time & 4.5~hours \\
        On-source time & 12.4~hours \\
        \hline
    \end{tabular}
\end{table}

%%% 

\subsection{Calibration}
\label{sec:calibration}

We have used the fully automated pipeline Source Peeling and Atmospheric Modeling (SPAM; \citealt{Intema2009, Intema2014a, Intema17}) to calibrate the GSB data used in this analysis. The SPAM pipeline has two major parts: a pre-processing and a main processing. In the pre-processing step, the observed data from each night is converted into a `pre-calibrated' UVFITS dataset for each PC. Next, for each PC, all the night's pre-calibrated data are fed into the main processing pipeline to obtain continuum images. We briefly summarize the steps below and refer the reader to \cite{Intema17} for a complete description.

In pre-processing, an initial round of flagging is performed on each scan of the calibrators, after which time-variable complex gain solutions and time-constant bandpass solutions per antenna and per polarization are computed. These steps (flagging, gain, and bandpass calibration) are repeated several times, applying increasingly strict flagging to obtain improved calibration results. After applying the bandpass and time-averaged gain solutions to the calibrator data, the RR and LL polarizations are combined to Stokes I, and data are averaged in frequency and time. The final number of channels after averaging is 42 of width 0.39~MHz, which yields an effective bandwidth of 16.7~MHz. The same calibration and averaging are performed on all seven target PCs.  

After pre-processing, the $\textit{uv}$-data from all three nights are combined and fed into the main pipeline, which is run on each PC individually. For each PC, the pipeline performs a direction-independent (DI) self-calibration, followed by a direction-dependent (DD) ionospheric calibration. Considering the DI self-calibration, 3 rounds of phase-only self-calibration are followed by one round of amplitude and phase self-calibration. Also, note that each round of the DI self-calibration is followed by wide-field imaging and CLEAN deconvolution of the primary beam out to 5 primary beam radii to include bright outlier sources to avoid the sidelobes of those sources during imaging. After the DI calibration steps, DD gain phases are obtained by peeling bright sources in the FoV, and this yields measures of ionospheric phase delay. DD gain phases per time stamp are fitted with a two-layer phase screen model. During imaging of the full FoV, this model is used to calculate the phase correction while applying the DD gain tables on the fly. The DD calibration requires at least four peeled sources for ionospheric modeling. The pipeline fails when fewer than four sources are found due to limitations in dynamic range (refer to Section 3.2.3 of \citealt{Intema17}). We found that the DD calibration has been successfully processed for all PCs from the observations, except for one night (17 Nov) of the observation of PC~A. We flagged this night for PC~A to avoid including any artifacts in the final image. The flagging percentage for PC~A is therefore slightly higher $(\approx54\%)$ compared to the rest of the PCs $(\sim30\%)$ (Table~\ref{table:ra_dec}). 

The end products of the SPAM pipeline are the calibrated and residual (source-subtracted) Stokes~I data, and the images before and after primary beam correction, for each PC. Note that we have restricted the extent of the images till the primary beam response falls to $30\%$ of its peak value, which corresponds to a radius of $\sim2^{\circ}$. Note that the FWHM of uGMRT at the reference frequency of $147.4$ MHz is $\sim3^\circ$. We have used the primary-beam corrected images for the rest of the work presented in this paper.  

\section{Images}
\label{sec:images}

In this section, we present the images produced by SPAM and how we combined these to make a single mosaic. We note that we have used Python Blob Detection and Source Finder (\textsc{PyBDSF}; \citealt{PyBDSF}) to estimate the background rms maps of all the images shown in this paper. Unless explicitly mentioned, the median of the rms map is quoted as the rms of the image. The details of the parameter choices in \textsc{PyBDSF} are given in Appendix~\ref{sec:pybdsf}.

% Individual PCs. 
\begin{figure*}
    \includegraphics[width=\textwidth]{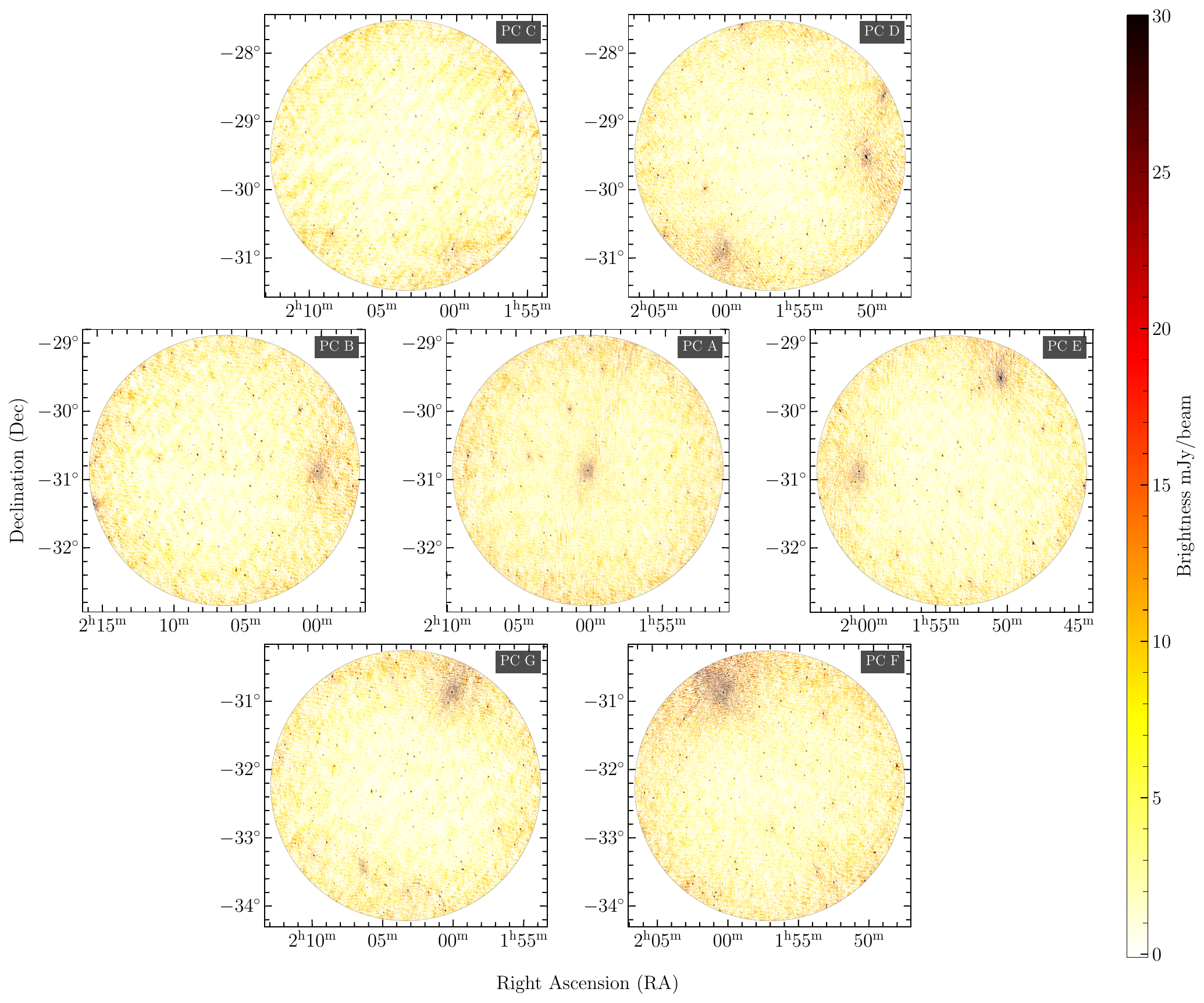}
    \caption{The primary beam corrected images of all the PCs. The extent of the images is restricted to 30\% of the peak primary beam response. The coordinates of each PC, the rms of the images, and the PSFs are given in Table~\ref{table:ra_dec}.}
    \label{fig:fits_plot}
\end{figure*}

Figure~\ref{fig:fits_plot} shows the primary beam corrected images for all the PCs.  Considering PC~A, we see the bright calibrator source GLEAM~02H at the centre of the image (Table~\ref{table:ra_dec}; PC~A). We notice some residual artifacts near this source and also around other brighter sources in this image. These artifacts are more prominent in the other PCs where this source appears at the sidelobe of the primary beam. We note that the observation duration per PC is relatively short ($\sim1.8$~hours), which further gets affected by significant data flagging due to RFIs. This results in a patchy $uv$-coverage, which leads to prominent sidelobe patterns in the PSF. The artifacts around the bright sources possibly arise due to improper deconvolution of the PSF sidelobes. A full synthesis image ($\sim8$~hours) may help achieve improvement in this aspect. The $uv$-coverage and the PSFs are further discussed in Appendix~\ref{sec:uvtracks}. Apart from the central bright source, we also notice another bright source at ${\rm RA, Dec} \approx 27.6^\circ, -29.5^\circ$ in PCs~D~and~E. A close inspection, however, shows that these are two sources. A detailed image of this source will require dedicated observation centring this ${\rm RA, Dec}$, which is planned for the uGMRT observation Cycle 49 (ID: $49\_019$).

Table~\ref{table:ra_dec} presents the rms values (before primary beam correction) and the PSFs of the images of individual PC. We find that the median rms values for different PCs vary in the range $2-3$~mJy/beam. PC~C seems to be the cleanest, whereas PC~A and F have higher rms values. Considering each PC, we find spatial variations in sensitivity across the image. The visible artifacts, as well as the rms, go up near the bright sources. We also find the artifacts and the rms to be higher toward the edges of the beam due to the effects of the primary beam correction. The median rms values after the primary beam correction vary in the range $4.5-5.9$~mJy/beam. 
 
We have combined the primary beam corrected images of all the PCs to create a single, deeper mosaic image. To achieve this, we first read the RA, Dec, and PSF information from all the primary beam-corrected images generated by SPAM. The FWHM of the PSFs of these images are found to differ slightly from one another (see Table~\ref{table:ra_dec}). We have used the CASA task \texttt{imsmooth} to match their PSFs. Next, we use the available RA, Dec values to determine the optimal celestial projection system that covers the entire field using the \texttt{find\_optimal\_celestial\_wcs} function from the \texttt{reproject} package of \textsc{Astropy} \citep{astropy13, astropy18}. We then reproject each image onto this common celestial grid using \texttt{reproject\_interp}. After reprojection, we co-added the images using an inverse-variance weighting scheme, where each pixel contributed proportionally to its signal-to-noise ratio. The inverse of the background rms images, which are obtained by using \textsc{PyBDSF} on each primary beam corrected image, are used as the weights here.

% Mosaic
\begin{figure*}
    \includegraphics[width=\textwidth]{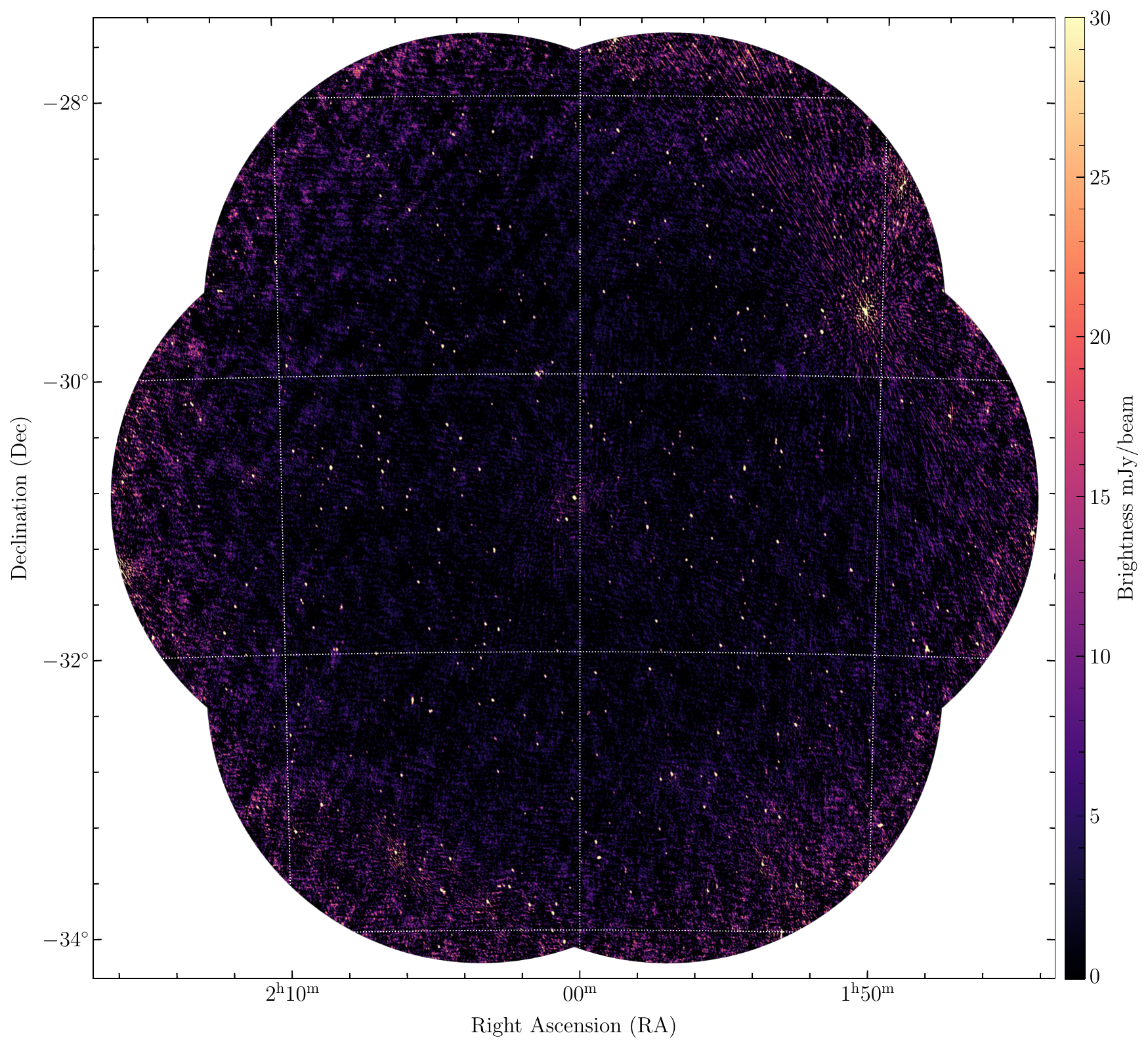}
    \caption{The final mosaic image of the HERA calibration field GLEAM~02H. It is made by combining the primary beam corrected images of all the 7 PCs, which are shown in Figure~\ref{fig:fits_plot}.}
    \label{fig:mosaic_plot}
\end{figure*}

% zoomed image. 

\begin{figure*}
    \includegraphics[width=0.95\textwidth]{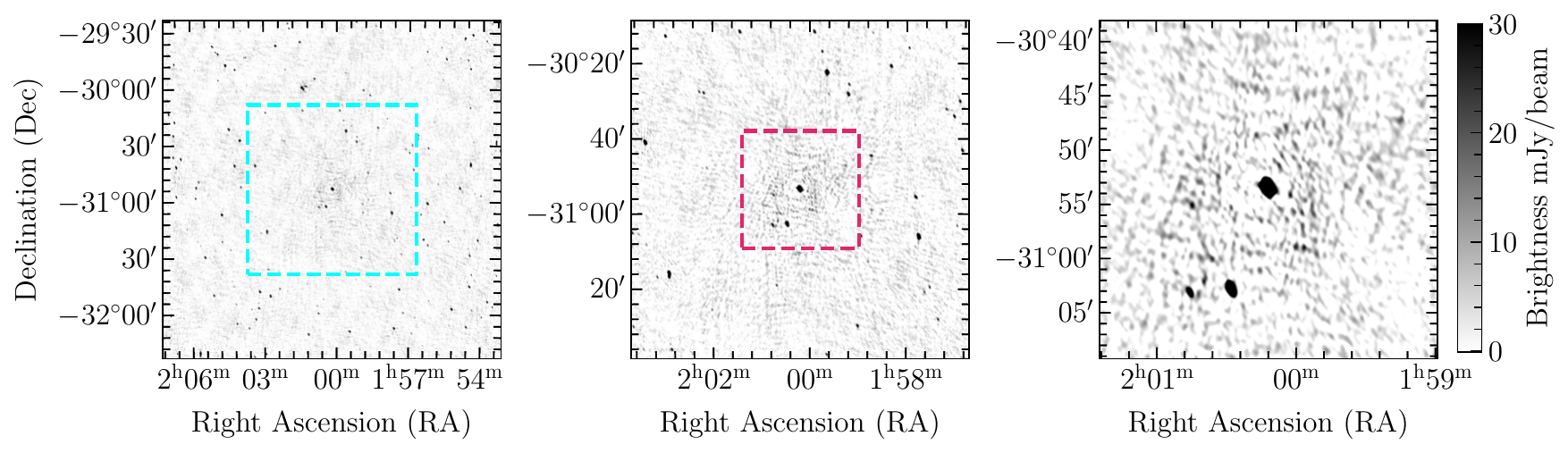}
    \caption{A zoomed-in view of the central region of the mosaic image. The left panel shows the central $3^\circ \times 3^\circ$ region, where we expect relatively uniform sensitivity. The cyan rectangle outlines a $1.5^\circ \times 1.5^\circ$ region, which is shown in more detail in the middle panel. The magenta square in the middle panel highlights a $0.5^\circ \times 0.5^\circ$ region, which is further magnified in the rightmost panel. The bright source at the center of these zoomed regions is the  Quasar PKS~0157--31.}
    \label{fig:mosaic_plot_triple}
\end{figure*}

Figure~\ref{fig:mosaic_plot} presents the mosaic image, which covers $6.9^\circ \times 6.9^\circ$ on the sky. We can notice several bright ($>1$~Jy) sources in the image, which were difficult to visualize in Figure~\ref{fig:fits_plot}. The artifacts around the central bright source and other sources have also been reduced to some extent. The central part of the mosaic has higher sensitivity as multiple PCs have overlapped in this region. The mosaic has less sensitivity in the periphery region, where the rms is worse due to (a) no overlaps between PCs, and (b) the primary beam correction in the images of the individual PCs.  To estimate the dynamic range of the image, we use \texttt{casaviewer} to interactively search for source-free regions in the image. We have chosen a square box of size $150 \times 150$ pixels (i.e., $10'$) at multiple source-free regions and found the rms to be $\sigma_{\rm rms}\sim2$~mJy/beam. The maximum brightness ($S_{{\rm peak}}$) of the mosaic is found to be $14.31$~Jy/beam, thereby yielding a dynamic range of $S_{{\rm peak}}/\sigma_{\rm rms} \approx 6500$.  A summary of the mosaic image is listed in Table~\ref{tab:mosaic_summary}.  

\begin{table}
    \centering
    \renewcommand{\arraystretch}{1.4} % Adjust row height
    \setlength{\tabcolsep}{4pt} % Adjust column spacing
    \caption{Summary of the mosaic image.}
    \label{tab:mosaic_summary}
    \begin{tabular}{l c} 
        \hline
        Parameter & Value \\
        \hline
        Pixel size & $4.10 \times 4.10$ arcsec \\
        Image size & $6077 \times 6074$ pixels ($6.9^\circ \times 6.9^\circ$) \\
        rms (median) & $3.9^{+3.7}_{-1.4}$~mJy/beam \\
        Off-source noise ($\sigma_{\rm{rms}}$) & $\sim2$~mJy/beam \\
        Maximum brightness ($S_{\rm{peak}}$) & $14.31$~Jy/beam \\
        Minimum brightness ($S_{\rm{min}}$) & $-0.13$~Jy/beam \\
        Dynamic range ($S_{\rm{peak}} / \sigma_{\rm{rms}}$) & $\sim6500$ \\
        FWHM of PSF (Maj $\times$ Min, PA) & $40'' \times 22'', 23^\circ$ \\
        \hline
    \end{tabular}
\end{table}

% rms map & % completeness
\begin{figure*}
    \centering
    \includegraphics[width=0.95\textwidth]{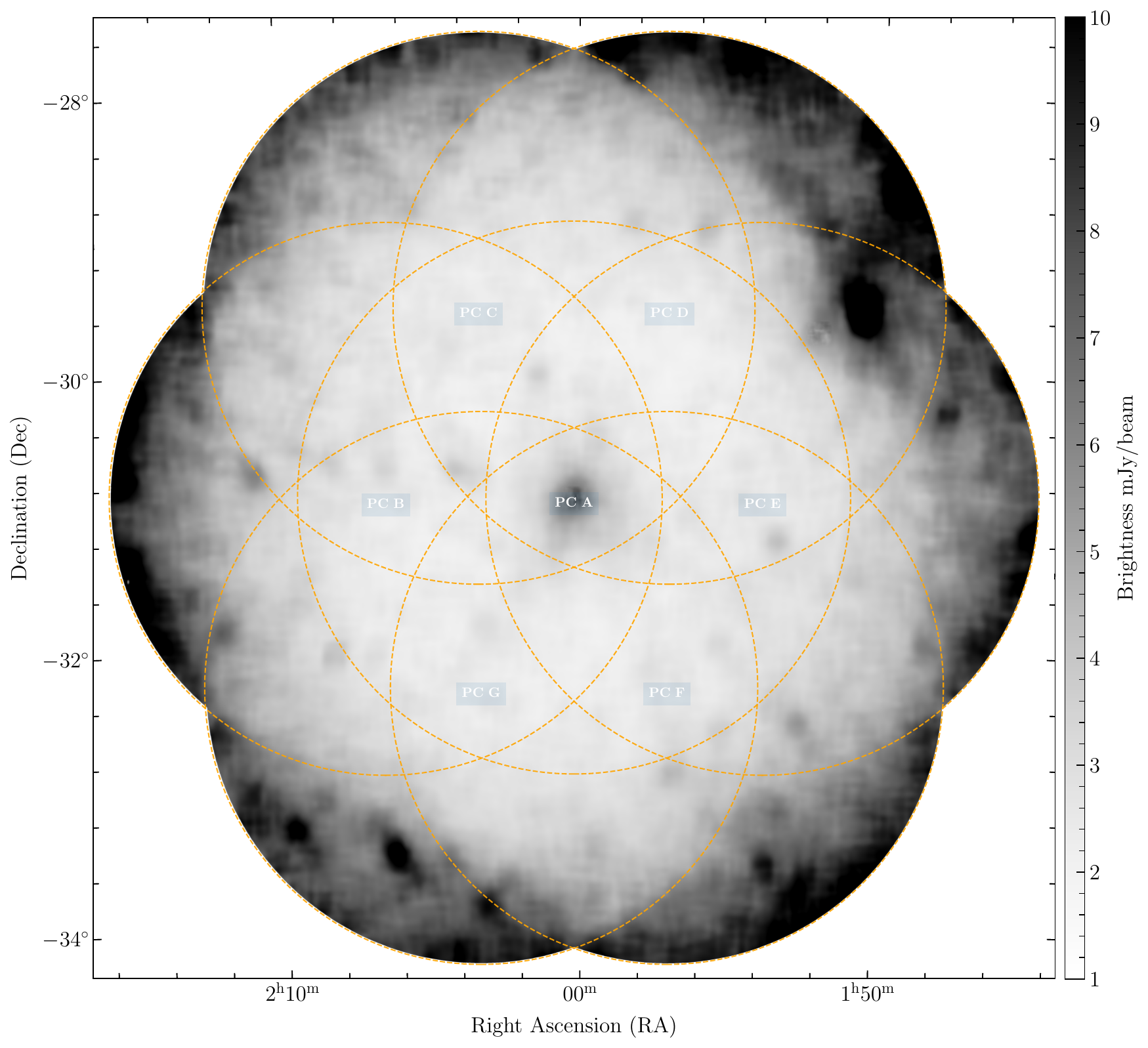}
    \caption{The rms map of the mosaic. The dashed circles, each of a radius of $2^\circ$, represent the area covered by the individual PCs, annotated at the centre of each circle. The rms in the image varies from $1.9-54.6$~mJy/beam, but here the higher end of the colorbar is saturated at 10~mJy/beam. The rms is relatively higher near the bright sources and at the edges of the FoV of the individual PCs.}
    \label{fig:rmsmap}
\end{figure*}

% Brightest source in the image
Figure~\ref{fig:mosaic_plot_triple} shows a zoomed-in view of the central regions of the mosaic image. The left panel shows the central $3^\circ \times 3^\circ$ region of the mosaic where multiple PCs overlap and we obtain a near-uniform sensitivity. The region outside this region has visible artifacts from individual PCs shown in Figure~\ref{fig:fits_plot}. This central $3^\circ \times 3^\circ$ region is further zoomed in to the next two panels. 
The middle panel shows the central $1.5^\circ \times 1.5^\circ$ region of the mosaic. We can notice that the artifacts from the central bright source have brightness $\gtrsim5$~mJy/beam in most parts of this region. The artifacts become more prominent as we further zoom into the central $0.5^\circ \times 0.5^\circ$ region of the mosaic in the right panel, where the artifacts around the bright source have brightness $\gtrsim10$~mJy/beam. 
The brightest source in the GLEAM~02H field is in the middle of all the panels. This source is the Quasar PKS~0157--31 (from NASA/IPAC Extragalactic Database\footnote{\url{https://ned.ipac.caltech.edu/byname?objname=PKS+0157-31}}). This source is classified as an FR~II Quasar and was labelled OC--397 by \cite{Morganti1993}, who reported its peak flux density as 0.3~Jy at 6~cm using VLA observations (see their Figure~1). A further discussion on this source, along with the other brightest sources in the mosaic, is presented in Section~\ref{sec:bright}. A detailed morphology of this source is beyond the scope of this work, but might be an interesting subject in the context of HERA calibration. 

\begin{figure}
    \centering
    \includegraphics[width=\columnwidth]{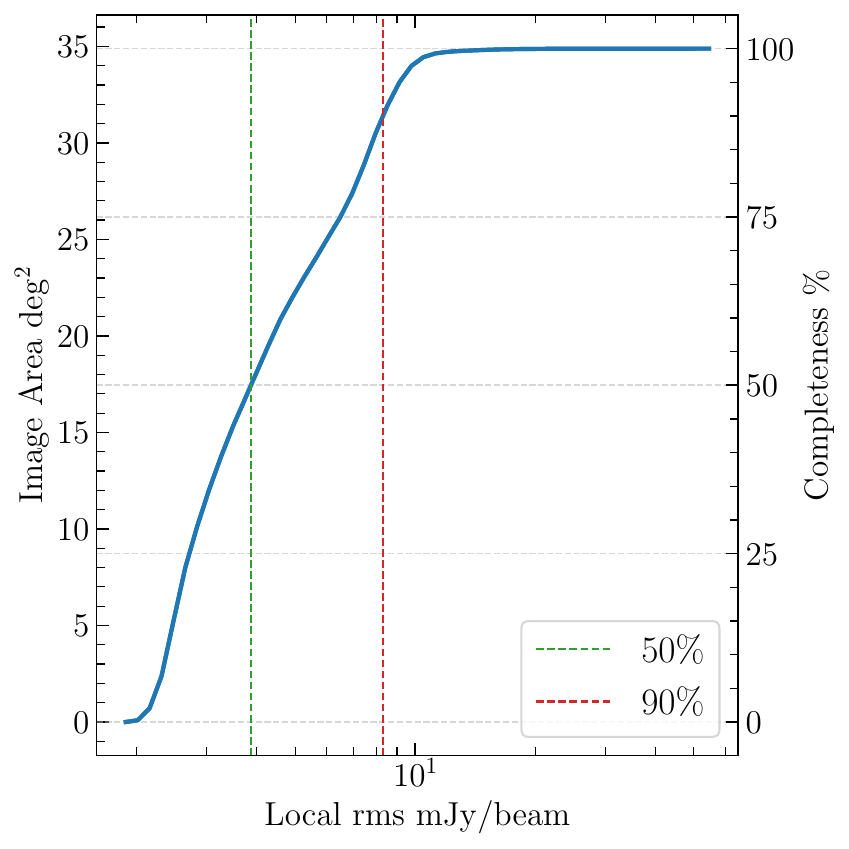}
    \caption{The cumulative area of the mosaic where the rms noise is less than or equal to the corresponding rms level on the x-axis. The right y-axis shows the same cumulative area expressed as a percentage of the total image area. The dashed vertical lines indicate 50\% and 90\% completeness levels of the mosaic.}
    \label{fig:completeness}
\end{figure}

Figure~\ref{fig:rmsmap} shows the rms map of the mosaic derived from \textsc{PyBDSF}. The dashed circles, each with a radius of $3^\circ$, represent the area covered by the individual PCs. The rms is relatively higher near the bright sources and at the edges of the FoV of each PC. This pattern in the rms appears due to the overlap of the PCs, which is due to the observation strategy that we follow (Figure~\ref{fig:observation_strategy}). We find the rms in the image to vary in the range from $\approx1.9 - 54.6$~mJy/beam, and the median value of the rms is found to be $3.9^{+3.7}_{-1.4}$~mJy/beam at the reference frequency of 147.4~MHz. Note that in the overlapping patch of the sky, but at a higher frequency of 200~MHz, GLEAM reports an rms sensitivity of $10 \pm 5$~mJy/beam (see Table~4 of \citealt{HW22}). 

We have used the rms map to calculate the cumulative area of the mosaic ($A_{\rm cum}$) that has a flux higher than a certain rms value. Figure~\ref{fig:completeness} shows $A_{\rm cum}$ as a function of the rms level. The figure also shows the cumulative area expressed as a percentage of the total image area ($A_{\rm {total}}$), i.e., $100\times A_{\rm cum}/A_{\rm {total}}$ on the right y-axis. We find that $50\%\,{\rm and} \, 90\%$ of the image area are above 3.9 and 8.3~mJy/beam, respectively. Since we have set a source detection limit of $5\sigma$, we are able to detect all the sources above 19.4 and 41.5~mJy/beam, respectively, from more than $50\%\,{\rm and} \, 90\%$ of the image area. In other words, the image is $50\%$ complete at 19.4~mJy and $90\%$ complete at 41.5~mJy. We note that the GLEAM catalogue in this part of the sky is estimated to be $50\%$ complete at $\sim50$~mJy and $\sim90\%$ complete at $100$~mJy at 200~MHz. GLEAM--X has improved upon GLEAM, and it is estimated to be $50\%$ complete at $\sim5.8$~mJy and $\sim90\%$ complete at $10.2$~mJy at 200~MHz. 

\section{Source Catalogue}
\label{sec:catalog}

We have used \textsc{PyBDSF} to find the sources from the mosaic and make a catalogue. As mentioned before, the parameters used in \textsc{PyBDSF} are described in Appendix~\ref{sec:pybdsf}. The catalogue comprises 640 sources in the flux density range  10.1~mJy to 19.1~Jy. Table~\ref{tab:catalogue} shows a sample of the catalogue. The full catalogue with all 45 columns is available online, with the descriptions provided in Appendix~\ref{sec:catalog_tab_def}.

\begin{table*}
\centering
\caption{A sample of the source catalogue. The columns show source ID, RA and Dec, total and peak flux densities and their errors, major (Maj) and minor (Min) axes, and position angle (PA) of the FWHM of the Gaussian fits, and island rms noise level. All the values are rounded to the second decimal place. The full table with all 45 columns is available online, with the descriptions provided in Appendix~\ref{sec:catalog_tab_def}.}
\label{tab:catalogue}
\begin{tabular}{ccccccccccc}
\hline
Source\_id & $\mathrm{RA}$ & $\mathrm{Dec}$ & $\mathrm{Total\_flux}$ & $\mathrm{E\_Total\_flux}$ & $\mathrm{Peak\_flux}$ & $\mathrm{E\_Peak\_flux}$ & $\mathrm{Maj}$ & $\mathrm{Min}$ & $\mathrm{PA}$ & $\mathrm{Isl\_rms}$ \\
 & (deg) & (deg) & (Jy) & (Jy) & (Jy/beam) & (Jy/beam) & (arcsec) & (arcsec) & (deg) & (mJy/beam) \\
\hline
0 & 33.91 & -30.80 & 0.11 & 0.03 & 0.07 & 0.01 & 47.62 & 29.60 & 47.13 & 10.68 \\
1 & 33.86 & -31.36 & 4.07 & 0.05 & 3.31 & 0.02 & 39.95 & 26.93 & 19.88 & 24.07 \\
2 & 33.83 & -30.73 & 0.75 & 0.03 & 0.55 & 0.01 & 42.82 & 27.66 & 24.75 & 12.16 \\
3 & 33.79 & -30.42 & 0.27 & 0.03 & 0.12 & 0.01 & 55.97 & 33.83 & 17.91 & 10.40 \\
\hline
\end{tabular}
\end{table*}

\subsection{Source Classification}
\label{sec:classification}

For the purpose of absolute calibration of any interferometer, e.g., HERA, it is essential to model bright sources with high fidelity. A key aspect of this is to determine whether a source is point-like or extended at the resolution of our uGMRT observations. Point-like sources can be accurately represented as delta functions in sky models, therefore it is straightforward to model the flux and spectral information. Extended sources, on the other hand, may exhibit spatial structure that introduces modeling errors if not properly accounted for. However, we note that a detailed investigation on the impact of source morphology on HERA absolute calibration is beyond the scope of this work.

In order to classify the sources in our catalogue as point sources or extended structures, we use a statistical approach based on the ratio of the integrated flux density $S_{\rm int}$ to the peak flux density $S_{\rm peak}$. For a point source, the ratio $S_{\rm int} /S_{\rm peak}$ is expected to be close to unity. To include the measurement uncertainties in this ratio, we define the logarithmic ratio as \citep{Franzen2015, Franzen2019}
    \begin{equation}
        R = \ln \left( \frac{S_{\rm int}}{S_{\rm peak}} \right)\,.
    \end{equation}
We obtain the uncertainty in $R$ using, 
    \begin{align}
        \sigma_R^2 &= \left( \frac{\partial R}{\partial S_{\rm int}} \sigma_{S_{\rm int}} \right)^2 + \left( \frac{\partial R}{\partial S_{\rm peak}} \sigma_{S_{\rm peak}} \right)^2 \\
        &= \left( \frac{\sigma_{S_{\rm int}}}{S_{\rm int}} \right)^2 + \left( \frac{\sigma_{S_{\rm peak}}}{S_{\rm peak}} \right)^2.
    \end{align}
where, $\sigma_{S_{\rm int}}$ and $\sigma_{S_{\rm peak}}$, the uncertainties in $S_{\rm int}$ and $S_{\rm peak}$, respectively, are independent and are added in quadrature. We classify the sources as point sources if $|R| \leq 2\sigma_R$ and as an extended structure if $|R| > 2\sigma_R$. Since a negative value of $R$ for which  $|R|>2\sigma_R$ can be satisfied due to noise fluctuations, or calibration errors, and does not imply an actual extended source, the condition $|R| \leq 2\sigma_R$ ensures that there is only a 2.3\% chance of an incorrect classification \citep{Franzen2015, Franzen2019}.

\begin{figure}
    \includegraphics[width=\columnwidth]{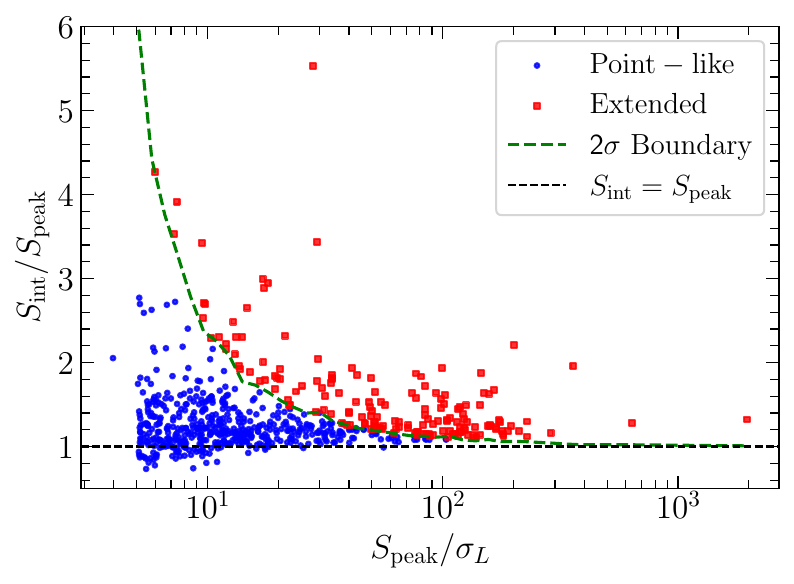}
    \caption{The ratio of integrated to peak flux density $S_{\rm int} /S_{\rm peak}$ as a function of signal-to-noise ratio $S_{\rm peak} /\sigma_L$  (where $\sigma_L$ is the local rms noise estimated by PyBDSF) of sources. The blue circles and red squares show point-like and extended sources, respectively. The green dashed line shows the $2\sigma$ boundary that demarcates the point-like and extended sources.}
    \label{fig:flux_ratio_vs_snr}
\end{figure}

Figure~\ref{fig:flux_ratio_vs_snr} shows the ratio of integrated to peak flux density $S_{\rm int} /S_{\rm peak}$ as a function of signal-to-noise ratio $S_{\rm peak} /\sigma_L$ of sources, where $\sigma_L$ is the local rms. The blue circles and red squares show point-like and extended sources, respectively.
We find that 164 out of 640 sources ($\approx26$\%) are classified as extended, while the remaining are point-like. We find that the brightest sources are more likely to be extended, while a significant number of the faint sources appear to be point-like.

\subsection{Cross-matching}
\label{sec:crossmatch}

We have assessed the astrometric and integrated flux density measurement accuracy of our catalogue by cross-matching it with the following overlapping radio source catalogues: TGSS\footnote{\url{https://cdsarc.cds.unistra.fr/viz-bin/cat/J/A+A/598/A78}} \citep{Intema17}, NRAO VLA Sky Survey (NVSS\footnote{\url{https://cdsarc.cds.unistra.fr/viz-bin/cat/VIII/65}}; \citealt{Condon1998}), GLEAM\footnote{\url{https://cdsarc.cds.unistra.fr/viz-bin/cat/VIII/100}} \citep{HW17}, and GLEAM--X DRII\footnote{\url{https://cdsarc.cds.unistra.fr/viz-bin/cat/VIII/113}} \citep{Ross2024}. The cross-matching is performed using a match radius of $30''$, which is $75\%$ of the FWHM of the PSF of our mosaic ($40''$). We have matched with the nearest sources between the catalogues. 

\begin{figure}
    \includegraphics[width=\columnwidth]{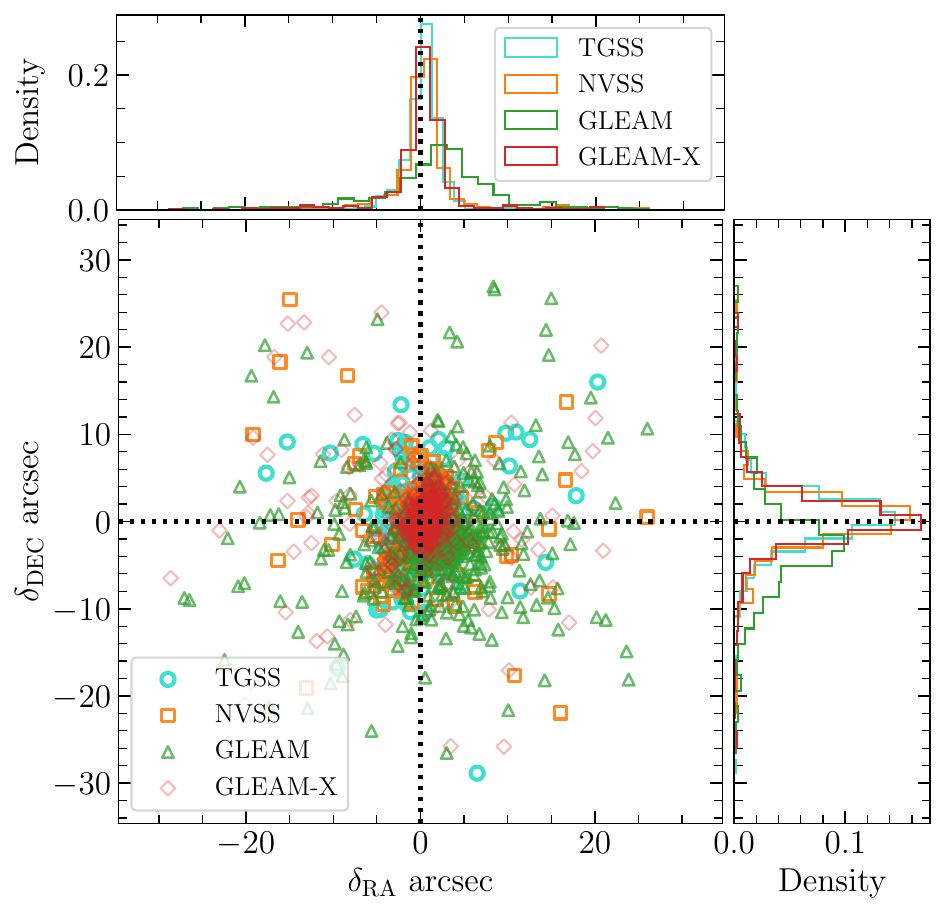}
    \caption{The positional offsets between uGMRT sources and their cross-matched counterparts in external catalogues. The small median offsets and tight distributions for TGSS, NVSS, and GLEAM--X reflect good astrometric alignment. The offsets are slightly larger for GLEAM, possibly due to its lower resolution.}
    \label{fig:positional_offsets}
\end{figure}

We have used the cross-matched sources in the uGMRT and the external catalogues to calculate their positional offsets on the sky. The positional offsets along the RA ($\delta_{\rm RA}$) and Dec ($\delta_{\rm Dec}$) are defined as,
\begin{align}
    \delta_{\rm RA} &= (RA_{\rm uGMRT} - RA_{\rm ext} ) \nonumber \cos{(\rm{Dec}_{\rm uGMRT})}\\
    \delta_{\rm Dec} &= {\rm Dec}_{\rm uGMRT} - {\rm Dec}_{\rm ext} \,.
\end{align}
Figure~\ref{fig:positional_offsets} shows the positional offsets $(\delta_{\rm RA}, \delta_{\rm Dec})$ for the cross-matched sources for the different catalogues. The median of the positional offsets and their uncertainties are summarized in Table~\ref{tab:crossmatch}. The median offsets with TGSS, NVSS, and GLEAM--X remain within $\approx0.5$~arcsec, which indicates negligible systematic deviations and sub-arcsecond astrometric accuracy. In contrast, we find comparatively larger offsets ($\delta_{\rm RA} = 2.35^{+2.72}_{-3.63}$~arcsec and $\delta_{\rm Dec} = -2.66^{+3.08}_{-3.00}$~arcsec) with respect to GLEAM. This increased offset associated with GLEAM is likely attributable to its relatively lower angular resolution, which has been enhanced in the GLEAM--X catalogue \citep{HW22}. 
% We do not find any systematic positional offset with respect to TGSS, NVSS, and GLEAM--X.

\begin{table}
\centering
\renewcommand{\arraystretch}{1.4}
\setlength{\tabcolsep}{8pt}
\caption{Median positional offsets and flux scale ratios relative to external catalogues. 
The uncertainties represent the 25th--75th percentile range.}
\label{tab:crossmatch}
\begin{tabular}{lccc}
\hline
Catalogue & $\delta_{\rm RA}$ (arcsec) & $\delta_{\rm Dec}$ (arcsec) & $S_{\rm uGMRT} / S_{\rm ext}$ \\
\hline
TGSS & $0.45^{+0.92}_{-1.01}$ & $0.41^{+1.79}_{-1.95}$ & $0.91^{+0.14}_{-0.11}$ \\ [6pt]
NVSS & $0.41^{+0.95}_{-1.01}$ & $0.19^{+1.68}_{-1.86}$ & -- \\ [6pt]
GLEAM & $2.35^{+2.72}_{-3.63}$ & $-2.66^{+3.08}_{-3.00}$ & $0.92^{+0.16}_{-0.21}$ \\ [6pt]
GLEAM-X & $0.43^{+0.95}_{-1.13}$ & $0.27^{+1.69}_{-1.55}$ & $1.00^{+0.15}_{-0.16}$ \\ [6pt]
\hline
\end{tabular}
\end{table}

\begin{figure}
    \centering
    \includegraphics[width=\columnwidth]{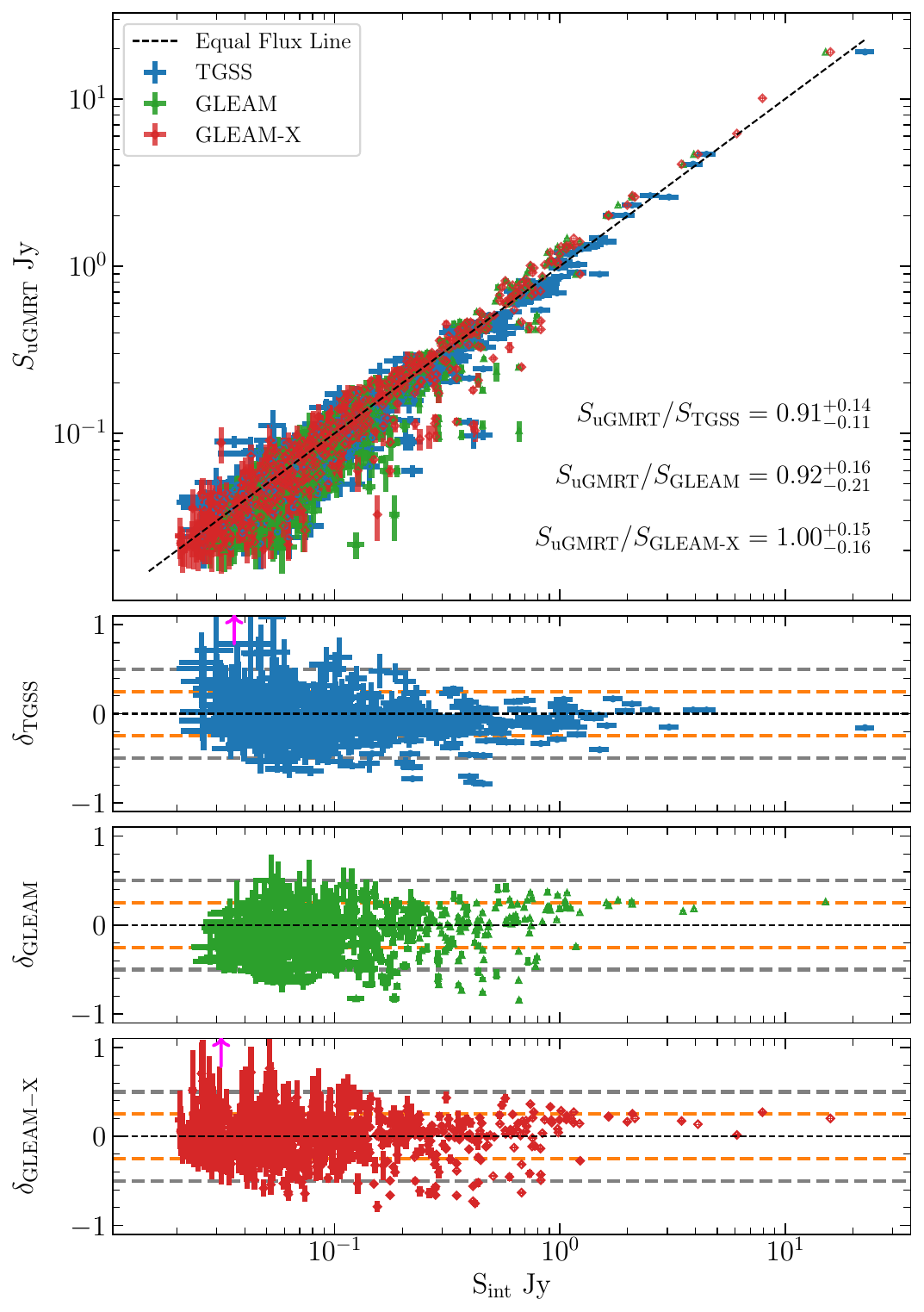}
    \caption{Comparison of uGMRT fluxes $S_{\rm uGMRT}$ with the fluxes of external catalogues $S_{\rm ext}$ for cross-matched sources. The median values and the corresponding uncertainties of the ratio $S_{\rm uGMRT} / S_{\rm ext}$ are annotated at the top panel (and also in Table~\ref{tab:crossmatch}). The bottom panels show the fractional deviation $\delta_{\rm ext} = (S_{\rm uGMRT} - S_{\rm ext}) / S_{\rm ext}$ with different catalogues. The horizontal dashed lines are drawn at $\delta_{\rm ext}$ = 0, 0.25, and 0.50, showing an exact match, $25\%$, and $50\%$ difference, respectively. Two matched sources for which $\delta_{\rm ext} > 1$ are shown with arrows.}
    \label{fig:flux_comparison}
\end{figure}

We also compare the integrated flux densities measured in our uGMRT catalogue with those from TGSS, GLEAM, and GLEAM--X catalogues. Considering TGSS, we have used the fluxes measured at 150 MHz, whereas we have used the fluxes measured at $151$~MHz for GLEAM and GLEAM--X. All the fluxes are extrapolated to $147.4$~MHz, the reference frequency of our catalogue, assuming the sources to have a power law with a spectral index of $-0.7$. Since all the flux measurements are at a very similar frequency, we do not expect the choice of the spectral index to play much of a role here. We note that the NVSS catalogue is compiled at a much higher frequency (1.4~GHz) than the rest, and so we exclude NVSS for the flux comparison. We will use the cross-matched NVSS sources for spectral index measurements (Section~\ref{sec:spindex}). Figure~\ref{fig:flux_comparison} shows the flux comparison along with the fractional deviation as a function of flux. 
The upper panel shows the extrapolated flux of the sources from other catalogues (TGSS in blue, GLEAM in green, and GLEAM-X in red) along the x-axis, whereas the y-axis shows the measured flux from this observation. The black dashed line shows the reference $45^{\circ}$ line, which indicates a match in the fluxes. 
% We see that the spread in the flux values is a bit large in lower flux regions, whereas the higher flux values match quite accurately with external catalogues. 
We see that the flux values match quite accurately for the higher fluxes, whereas there is a relatively larger spread in lower flux regions. 
The median flux ratios (uGMRT to external) and their uncertainties are annotated here and are also summarized in Table~\ref{tab:crossmatch}. The median flux ratio is close to unity for GLEAM--X, while these are $\approx 0.90$ for TGSS and GLEAM. 
The lower three panels show the fractional deviation $\delta_{\rm ext} = (S_{\rm uGMRT} - S_{\rm ext}) / S_{\rm ext}$ of the flux compared with those three external catalogues.  
Considering all three external catalogues, we see that the flux values match quite accurately for the sources with higher fluxes ($>0.8$~Jy), where we find $\delta_{\rm ext} < 25\%$ for most of the sources. Although the value of  $\delta_{\rm ext}$ becomes larger at lower fluxes, most of the sources are found to be within $\delta_{\rm ext} < 50\%$. 
We expect this difference in the lower flux region due to the different sensitivities and resolutions in different interferometers. In general, we find a good agreement in flux measurement between our catalogue and these catalogues within their calibration uncertainties. We conclude from the comparison of our uGMRT catalogue with established catalogues that it has astrometric precision at the sub-arcsecond level, with flux calibration that aligns well with earlier surveys. As a result, our catalogue is appropriate for developing sky models necessary for HERA calibration.

\subsection{Spectral index distribution}
\label{sec:spindex}

\begin{figure}
    \centering
    \includegraphics[width=\columnwidth]{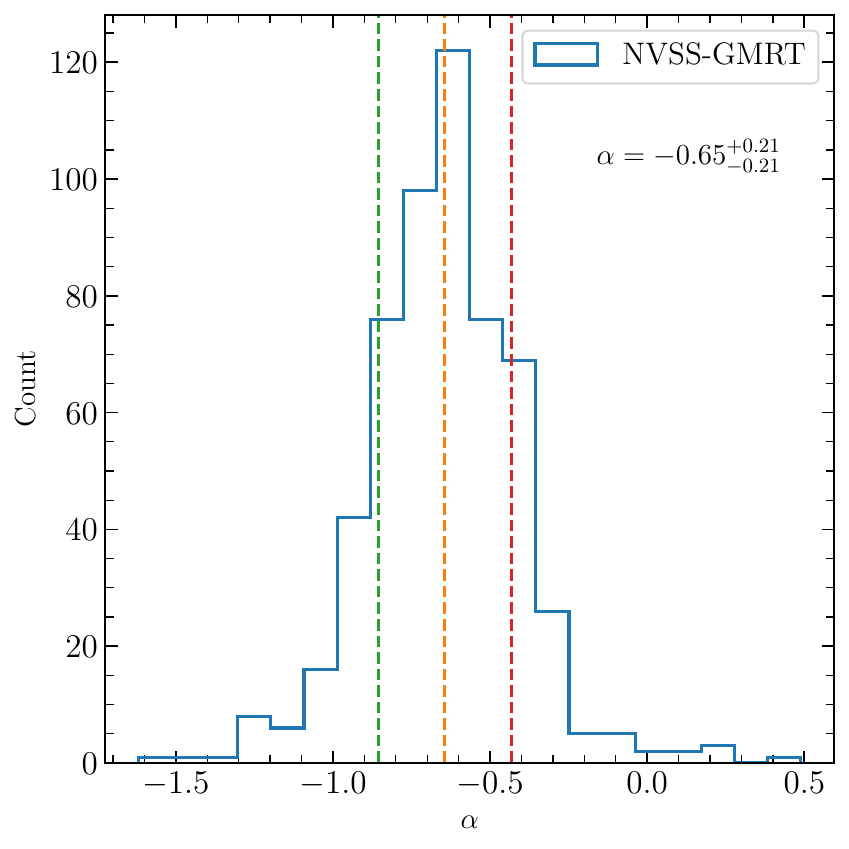}
    \caption{Distribution of spectral indices computed from sources cross-matched between the NVSS and our uGMRT catalogue.}
    \label{fig:spectral_index}
\end{figure}

We have computed spectral indices ($\alpha$) for sources cross-matched between our 147.4~MHz uGMRT catalogue and the 1.4~GHz NVSS catalogue. The sources are assumed to follow the standard relation $S_\nu \propto \nu^\alpha$, where $S_\nu$ is the flux density at frequency $\nu$. Figure~\ref{fig:spectral_index} shows the resulting distribution.
The spectral indices span the range $\alpha\approx -1.5$ to $0.5$, with a median value of $\alpha = -0.65^{+0.21}_{-0.21}$. We note that the median value is consistent with typical values for galactic synchrotron emission. 

% We note that the GWB data will provide 

\subsection{The brightest sources}
\label{sec:bright}

Table~\ref{tab:gmrt_gleam_nvss_alpha} presents the top 10 brightest sources from the uGMRT catalogue, sorted in descending order by total flux density, along with the cross-matched sources from other catalogues. 
%It is expected that these sources, particularly their spectral behaviour, will have the most important effect in the context of HERA's absolute calibration.
These sources are found to have integrated flux $>2$~Jy at 147.4~MHz, and therefore, these are the sources that dominate the overall flux of the images. The brightest source in our catalogue (ID 329) is associated with the GLEAM ID J020012--305324 in the GLEAM catalogue, and it is the main target source for the HERA absolute calibration. 
The measured integrated flux of this source is reported to be  17.95 and 18.86~Jy at 151~MHz in GLEAM and GLEAM--X catalogues, respectively. The spectral index of the same source is quoted as $\alpha_{\rm GLEAM} = -0.86 \pm 0.01$ and $\alpha_{\rm GLEAM-X} = -0.78 \pm 0.01$ in GLEAM and GLEAM--X, respectively. 
We use a power law $S_\nu \propto \nu^\alpha$ to scale the fluxes to our reference frequency of 147.4~MHz. We find that the scaled values of the flux at 147.4~MHz are 18.33 and 19.22~Jy for GLEAM and GLEAM--X catalogues, respectively. 
We report the flux of this source to be 19.08~Jy at 147.4~MHz, which is $\approx4\%$ higher than GLEAM, and it matches with GLEAM--X within $1\%$.
We further note that the TGSS quotes a slightly higher value for the flux of this source, but it has a larger uncertainty, and the measurement is consistent with these catalogues. We also note that a comparison with 1.4~GHz NVSS catalogue has yielded a spectral index $\alpha_{\rm NVSS} = -0.7$, which is slightly flatter than both $\alpha_{\rm GLEAM}$ and $\alpha_{\rm GLEAM-X}$. The GWB data might yield a better constraint on the spectral index of this source.  We note here that for all the other sources, $\alpha_{\rm NVSS}$ is found to be quite close to $\alpha_{\rm GLEAM}$ and $\alpha_{\rm GLEAM-X}$. 

\begin{table*}
    \centering
    \renewcommand{\arraystretch}{1.2} % Adjust row height
    \setlength{\tabcolsep}{4pt}
    \caption{The top 10 brightest sources from the uGMRT catalogue, sorted in descending order by total flux density, with the cross-matched sources from other catalogues. The columns show source IDs from our catalogue and GLEAM, the RA, Dec values in sexagesimal, and total flux densities with uncertainties at $\sim150$~MHz from uGMRT, GLEAM, GLEAM--X, and TGSS. The last three columns show the spectral indices quoted in the GLEAM, GLEAM--X, and those derived using our catalogue by comparing with the 1.4~GHz NVSS catalogue (Section~\ref{sec:spindex}).}
    \label{tab:gmrt_gleam_nvss_alpha}
    \begin{tabular}{cccccccccccc}
    \hline
    ID & GLEAM ID & RA & Dec & $S_{\rm int}$ & $S_{\rm int, GLEAM}$  & $S_{\rm int, GLEAM-X}$& $S_{\rm int, TGSS}$ & $\alpha_{\rm GLEAM}$ & $\alpha_{\rm GLEAM-X}$ & $\alpha_{\rm NVSS}$ \\
     &  & (hh:mm:ss) & (dd:mm:ss) & Jy & Jy & Jy & Jy &  &  &  \\
    \hline
    329  & J020012--305324 & 02:00:12 & --30:53:28 & $19.08 \pm 0.03$ & $17.95 \pm 0.02$ & $18.86 \pm 0.00$ & $22.31 \pm 2.23$ & $-0.86 \pm 0.01$ & $-0.78 \pm 0.01$ & $-0.72$ \\[3pt]
    584  & J015035--293158 & 01:50:34 & --29:32:37 & $10.08 \pm 0.12$ & $16.60 \pm 0.03$ & $9.36 \pm 0.00$  & $11.85 \pm 1.19$ & $-0.80 \pm 0.01$ & --               & $-0.79$ \\[3pt]
    585  & J015035--293158 & 01:50:38 & --29:31:11 & $6.20 \pm 0.08$  & $16.60 \pm 0.03$ & $7.39 \pm 0.00$  & $11.85 \pm 1.19$ & $-0.80 \pm 0.01$ & --               & $-0.70$ \\[3pt]
    558  & J015200--294056 & 01:52:01 & --29:41:03 & $4.67 \pm 0.02$  & $4.66 \pm 0.02$  & $4.92 \pm 0.00$  & $4.42 \pm 0.44$  & $-0.73 \pm 0.01$ & $-0.69 \pm 0.01$ & $-0.68$ \\[3pt]
    1    & J021527--312144 & 02:15:51 & --31:21:46 & $4.07 \pm 0.05$  & $3.95 \pm 0.02$  & $3.97 \pm -$     & $3.85 \pm 0.39$  & $-0.59 \pm 0.01$ & $-0.61 \pm 0.01$ & $-0.63$ \\[3pt]
    604  & J014930--283819 & 01:49:30 & --28:38:27 & $2.64 \pm 0.05$  & $2.48 \pm 0.02$  & $2.58 \pm 0.00$  & $2.48 \pm 0.25$  & $-0.89 \pm 0.02$ & $-0.82 \pm 0.01$ & $-0.80$ \\[3pt]
    89   & J020820--303928 & 02:08:24 & --30:39:34 & $2.59 \pm 0.01$  & $2.69 \pm 0.02$  & $2.94 \pm 0.00$  & $3.01 \pm 0.30$  & $-1.15 \pm 0.01$ & $-1.07 \pm 0.01$ & $-0.90$ \\[3pt]
    145  & J020620--332547 & 02:06:21 & --33:25:51 & $2.32 \pm 0.03$  & $2.14 \pm 0.02$  & $2.40 \pm 0.00$  & $2.06 \pm 0.21$  & $-0.76 \pm 0.02$ & $-0.73 \pm 0.01$ & $-0.76$ \\[3pt]
    238  & J020311--334722 & 02:03:12 & --33:47:25 & $2.01 \pm 0.02$  & $1.81 \pm 0.02$  & $1.88 \pm 0.00$  & $1.93 \pm 0.19$  & $-0.59 \pm 0.02$ & $-0.61 \pm 0.01$ & $-0.59$ \\[3pt]
    376  & J015853--273607 & 01:58:53 & --27:36:11 & $2.01 \pm 0.05$  & $1.89 \pm 0.02$  & $1.92 \pm 0.00$  & $1.70 \pm 0.17$  & $-0.68 \pm 0.02$ & $-0.68 \pm 0.01$ & $-0.68$ \\
    \hline
    \end{tabular}
\end{table*}

The next two brightest sources in the catalogue (ID 584 and 585) are identified as a single source in the GLEAM catalogue (GLEAM ID: J015035--293158). However, due to a relatively better angular resolution, these two sources are resolved in GLEAM--X.  While GLEAM provides a spectral index of $-0.80$ for this source, this information is not available for GLEAM--X. We note that this source appears at the edges of the FoV of the PCs D and E (Figure~\ref{fig:fits_plot}). A preliminary comparison with NVSS suggests that the spectral indices for the two sources are different. However, a detailed exposition on this source is deferred to future work that will involve the wideband data and larger sky coverage.

\section{Source Counts}
\label{sec:sourcecount}

We have estimated the differential source counts based on the flux densities obtained from our catalogue. The primary objective of this source count analysis is to perform a comparative assessment of our catalogue with GLEAM, specifically focusing on the flux densities achievable through this survey. However, we note that the study of the distribution of sources with flux density is important in its own right to understand the population of different radio galaxies, particularly at the low flux density regime. A direct quantification of source counts solely based on our catalogue does not show true extragalactic source distribution, particularly at the faint end of flux density bins. For example, the rms noise in the image varies in different regions, and we will detect the faint sources only within specific regions rather than across the entire image. This discrepancy leads to an underestimation of the source counts, thereby introducing a bias unless corrected appropriately. For the entire analysis presented in this section, we have used the flux density range $[S_i, S_f] = [20~{\rm mJy}, 19.1~{\rm Jy}]$, which includes 598 sources. We have divided the flux density range into 10 bins of equal logarithmic interval, and correct for false detection rate (FDR), incompleteness, Eddington bias, and visibility area effects, and estimate the differential source counts.

% We use simulations to correct for false detection rate (FDR), incompleteness, Eddington bias, and visibility area effects. For the entire analysis presented in this section, we have used the flux density range $[S_i, S_f] = [20~{\rm mJy}, 19.1~{\rm Jy}]$, and divided the range into 10 bins of equal logarithmic interval. The total number of sources in the flux range is 598.

\subsection{False Detection Rate (FDR) Correction}
\label{sec:fdr}

\begin{figure}
    \includegraphics[width=\columnwidth]{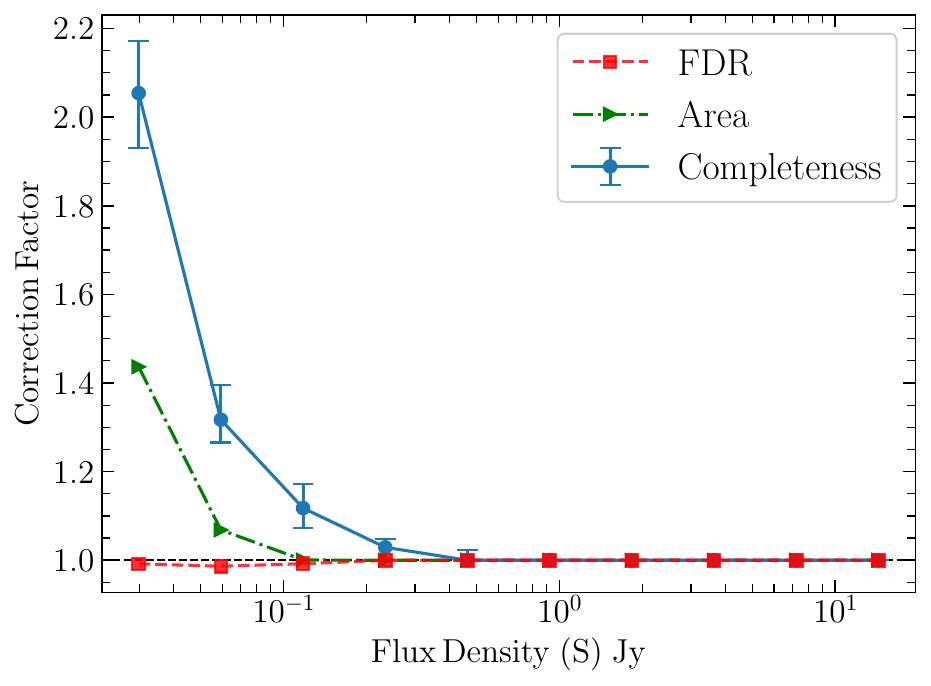}
    \caption{Corrections in the differential source count for false detection rate (FDR), visibility area, and incompleteness.}
    \label{fig:corrections}
\end{figure}

The source-finding algorithm \textsc{PyBDSF} can detect noise fluctuations and artifacts as sources. To account for these spurious source detections, we take the following standard procedure: We first invert the mosaic by multiplying the pixel values by $-1$. We run \textsc{PyBDSF} on the inverted image and count the number of detected sources as a function of flux density. The fraction of spurious detections in each flux bin is given by
    \begin{equation}
        f_{\rm{false}}(S) = \frac{N_{\rm{false}}(S)}{N_{\rm{total}}(S)},
    \end{equation}
where $N_{\rm{false}}(S) $ is the number of sources detected in the inverted image and $N_{\rm{total}}(S)$  is the number detected in the real image. The fraction of reliable sources in each flux bin is 
    \begin{equation}
        C_{\rm{fdr}}(S) = 1 - f_{\rm{false}}(S)\,.
    \end{equation}
Figure~\ref{fig:corrections} shows the correction factor $C_{\rm{fdr}}(S)$ of the sources for different flux density bins. We multiply the measured source counts by this correction factor $C_{\rm{fdr}}(S)$  to account for the false detections. The correction factors are also quoted in Table~\ref{tab:source_counts}. In total, we found only 4 spurious detections out of 598 sources, and we find that our detections have a reliability $100 \times C_{\rm{fdr}} > 98.6\%$ across the flux density range (Table~\ref{tab:source_counts}). 

\begin{table*}
    \centering
    \setlength{\tabcolsep}{10pt} % Adjust column spacing
    \renewcommand{\arraystretch}{1.5} % Adjust row spacing
    \caption{The Euclidean normalised differential source counts $S^{2.5}\,dN/dS$. The columns below correspond to the flux density bins $[S_i, S_f]$, the central value of the flux density bin $S$, bin widths ($\Delta S$), number of sources $N$, FDR correction ($C_{\rm fdr}$), area correction ($C_{\rm area}$), completeness correction ($C_{\rm c}$),  and Euclidean normalized source counts with errors. The Euclidean normalised source count is computed as: $ \left( \frac{N}{\Delta S \times A} \right) \times S^{2.5} \times C_{\rm fdr} \times C_{\rm area} \times C_{\rm c}$.} 
    
\begin{tabular}{ccccccccc}
    \hline
    $S_i$ & $S_f$ & $S$ & $\Delta S$ & N & $C_{\rm fdr}$ & $C_{\rm area}$ & $C_{\rm c}$ & $S^{2.5}\,dN/dS$ \\
    (Jy) & (Jy) &  (Jy) &  (Jy) &  &  &  &   &  (Jy$^{3/2}$ sr$^{-1}$) \\
    \hline
    0.020 & 0.040 & 0.030 & 0.0197 & 124 & $0.992$ & 1.436 & $2.054^{+0.117}_{-0.123}$ & $ 266.87 \pm 23.97 $ \\
    0.040 & 0.079 & 0.059 & 0.0392 & 148 & $0.986$ & 1.068 & $1.317^{+0.079}_{-0.051}$ & $ 422.85 \pm 34.76 $ \\
    0.079 & 0.157 & 0.118 & 0.0778 & 127 & $0.992$ & 1.001 & $1.118^{+0.054}_{-0.045}$ & $ 811.54 \pm 72.01 $ \\
    0.157 & 0.311 & 0.234 & 0.1544 & 101 & $1.000$ & 1.000 & $1.029^{+0.019}_{-0.029}$ & $ 1675.31 \pm 166.70 $ \\
    0.311 & 0.618 & 0.464 & 0.3067 & 49 & $1.000$ & 1.000 & $1.000^{+0.023}_{-0.000}$ & $ 2210.00 \pm 315.71 $ \\
    0.618 & 1.227 & 0.922 & 0.6090 & 32 & $1.000$ & 1.000 & $1.000^{+0.000}_{-0.000}$ & $ 4039.02 \pm 714.01 $ \\
    1.227 & 2.436 & 1.831 & 1.2094 & 10 & $1.000$ & 1.000 & $1.000^{+0.000}_{-0.000}$ & $ 3532.29 \pm 1117.01 $ \\
    2.436 & 4.838 & 3.637 & 2.4018 & 4 & $1.000$ & 1.000 & $1.000^{+0.000}_{-0.000}$ & $ 3954.10 \pm 1977.05 $ \\
    4.838 & 9.608 & 7.223 & 4.7696 & 1 & $1.000$ & 1.000 & $1.000^{+0.000}_{-0.000}$ & $ 2766.42 \pm 2766.42 $ \\
    9.608 & 19.079 & 14.343 & 9.4719 & 2 & $1.000$ & 1.000 & $1.000^{+0.000}_{-0.000}$ & $ 15483.81 \pm 10948.71 $ \\
    \hline
\end{tabular}

    \label{tab:source_counts}
\end{table*}

\subsection{Visibility Area Correction}

The effective area over which a source of flux density $S$ can be detected depends on the local noise level $\sigma_L$, which is not uniform over the image. The visibility function $V(S)$ accounts for this spatially varying sensitivity across the image. Without this correction, source counts can be biased because the faint sources at the detection limit are observable only in the most sensitive regions of the image. 

We have used the rms noise map of the mosaic (Figure~\ref{fig:rmsmap}) to calculate $A_{\rm eff}(S)$, the effective image area for each flux density bin. Only those pixels contribute to the effective area where the measured flux density is greater than the $5\sigma_L$ detection threshold. These pixel areas are then cumulatively added to calculate the effective area in each flux density bin.  In other words, for each flux density bin, a pixel contributes to the effective area if its detection threshold is lower than or equal to the lower edge of that bin and within the upper edge of the bin. This yields $A_{\rm eff}(S)$ the effective image area per flux density bin. We denote the visibility area correction factor as  $C_{\rm area}(S) = A/A_{\rm eff}(S)$, where $A$ is the total image area. The values of $C_{\rm area}(S)$ are presented in Table~\ref{tab:source_counts}, and are shown in Figure~\ref{fig:corrections}. The area correction is relatively higher at the lowest flux density bin that ranges between $20-40$~mJy. The area correction is much smaller at flux densities higher than $40$~mJy.

\subsection{Completeness Correction}
\label{sec:completeness_correction}

The catalogue is incomplete due to several reasons that need to be corrected. For example, the peak flux density of an extended source may be significantly reduced such that it goes below the local rms threshold. Although these extended sources have the same integrated flux density as the unresolved ones, we are unable to detect these sources and hence bias the source counts. This bias arises due to the resolution of the image and needs correction for an unbiased source count. It is also possible that noise fluctuation in the image can redistribute low flux density sources to higher flux bins. This is termed the Eddington bias and requires a correction. We use simulations prescribed in \cite{Williams2016, Hale2019, Franzen2019, Chakraborty2019b} to correct for these biases in the source counts. 

First, we generate 598 artificial sources from a power-law distribution \citep{Franzen2016}
    \begin{equation}
        \frac{dN}{dS} \propto S^{-1.6}
    \end{equation}
in the flux density range 20~mJy to 19~Jy, which is the same range used throughout this section. We add these sources to the residual image at random positions using the \textsc{AeRes} tool of the software \textsc{aegean}\footnote{\url{https://github.com/PaulHancock/Aegean}} \citep{Hancock2012aegean1, Hancock2018aegean2}. We have considered $26\%$ of these sources as extended sources with sizes greater than the PSF. From this image with injected sources, we have extracted the sources using \textsc{PyBDSF}, following the same detection criteria as described in Appendix ~\ref{sec:pybdsf}. The sources are then binned into the same 10 logarithmic flux density bins. The completeness correction factor for each flux density bin is computed as:
\begin{equation}
    C_{\rm c}(S) = \frac{N_{\mathrm{injected}}(S)}{N_{\mathrm{recovered}}(S)}
\end{equation}
where $N_{\mathrm{injected}}(S)$ and $N_{\mathrm{recovered}}(S)$ are the number of injected and recovered sources, respectively. We have performed 100 such simulations, from which the median value of  $C_{\rm c}(S)$ is chosen as the correction factor, while the 16th and 84th percentiles are its error bars. This method inherently accounts for resolution and Eddington bias \citep{Williams2016, Hale2019, Franzen2019, Chakraborty2019b}. The completeness correction factors and their error bars are shown in Figure~\ref{fig:corrections}, and also quoted in Table~\ref{tab:source_counts}.

\subsection{Euclidean normalized differential source counts}

\begin{figure*}
    \includegraphics[width=0.95\textwidth]{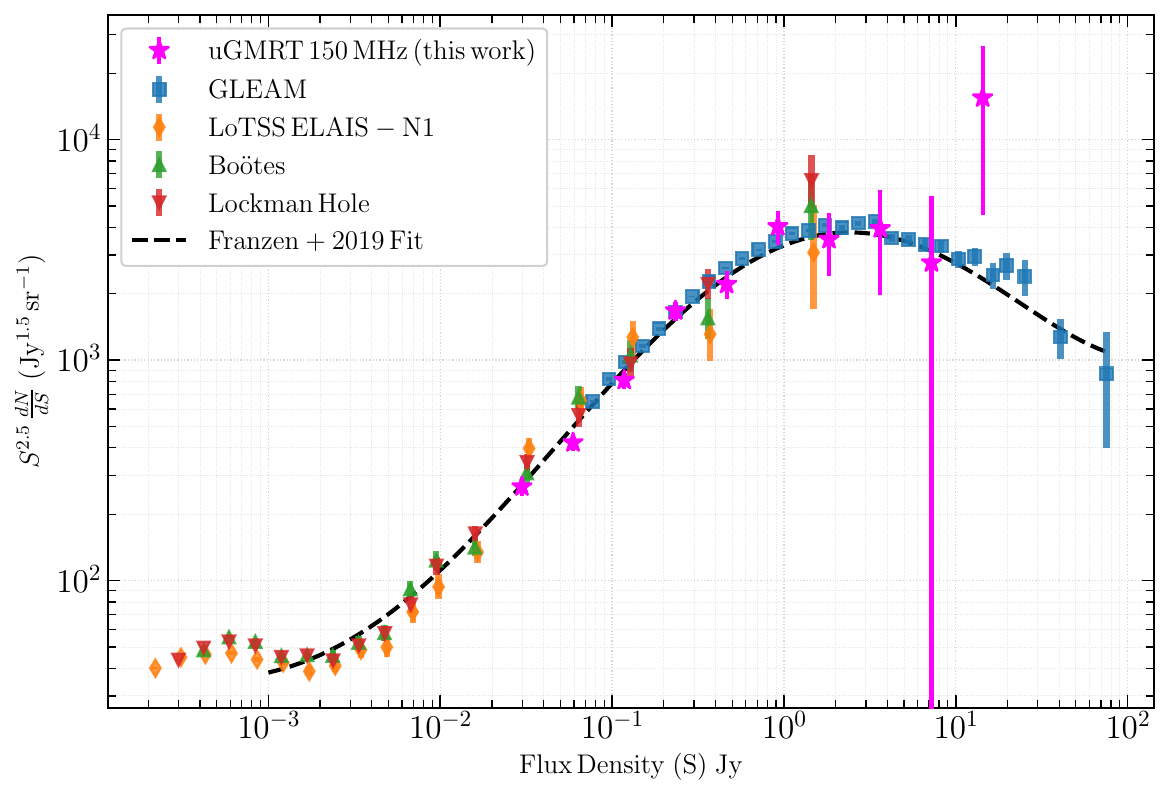}
    \caption{The Euclidean normalized differential source counts $S^{2.5}\,dN/dS$ with $1\sigma$ uncertainties as a function of flux densities. This also shows the source counts from the entire GLEAM survey and their best-fit polynomial \citep{Franzen2019}. Source counts from three LoTSS surveys (ELAIS-N1, Bootes, Lockman Hole) are also included to show the source counts at lower flux ranges.}
    \label{fig:source_counts_comparison}
\end{figure*}

The Euclidean normalised differential source count $S^{2.5}\,dN/dS$ is computed in each flux density bin using
\begin{equation}
    S^{2.5}\,dN/dS \equiv S^{2.5} \times \left( \frac{N}{\Delta S \times A} \right)  \times C_{\rm fdr} \times C_{\rm area} \times C_{\rm c} \, ,
    \label{eq:endsc}
\end{equation}
where $S$ is the central value of the flux density bin, $\Delta S$ is the bin width, and $A$ is the total image area. Here, for each flux density bin, $C_{\rm fdr}$, $C_{\rm area}$, and $C_{\rm c}$  account for the corrections due to FDR, visibility area, and completeness, respectively. 
Figure~\ref{fig:source_counts_comparison} shows $S^{2.5}\,dN/dS$ and their uncertainties as a function of flux densities $S$. This also shows the source counts from the GLEAM survey and their best-fit polynomial \citep{Franzen2019}. 
\begin{equation}
    \log_{10}\left(S^{2.5} \frac{dN}{dS}\right) = \sum_{i=0}^{5} a_{i} [\log_{10}(S)]^{i} \mathrm{,}
\end{equation}
where $a_{0} = 3.52$, $a_{1} = 0.307$, $a_{2} = - 0.388$, $a_{3} = -0.0404$, $a_{4} = 0.0351$ and $a_{5} = 0.00600$ \citep{Franzen2019}.
The figure also includes the source counts from the LOFAR Two Meter Sky Survey (LoTSS) of the fields ELAIS-N1, Bootes, and Lockman Hole \citep{Mandal2021} to compare the source counts at lower flux ranges. We find that the source counts are in line with GLEAM in the flux range $>100$~mJy. Note that there are only a total of 7 sources above $2.4$~Jy, and the error bars are significantly larger than GLEAM. Considering the flux densities $<100$~mJy, our source counts seem to closely follow both the LOFAR-based source counts and the model of \citep{Franzen2019}. Our catalogue extends the source counts down to $20$~mJy, probing flux densities more than a factor of three fainter than GLEAM, which reaches $69$~mJy. To our knowledge, these represent the deepest source counts yet measured in the GLEAM~02H field at $\sim150$~MHz. Such deeper catalogues are important for achieving complete sky coverage of HERA calibration fields and for accurate modelling of faint sources.

\section{Sky-model Simulations for HERA}
\label{sec:herasim}

\begin{figure*}
    \includegraphics[width=0.9\textwidth]{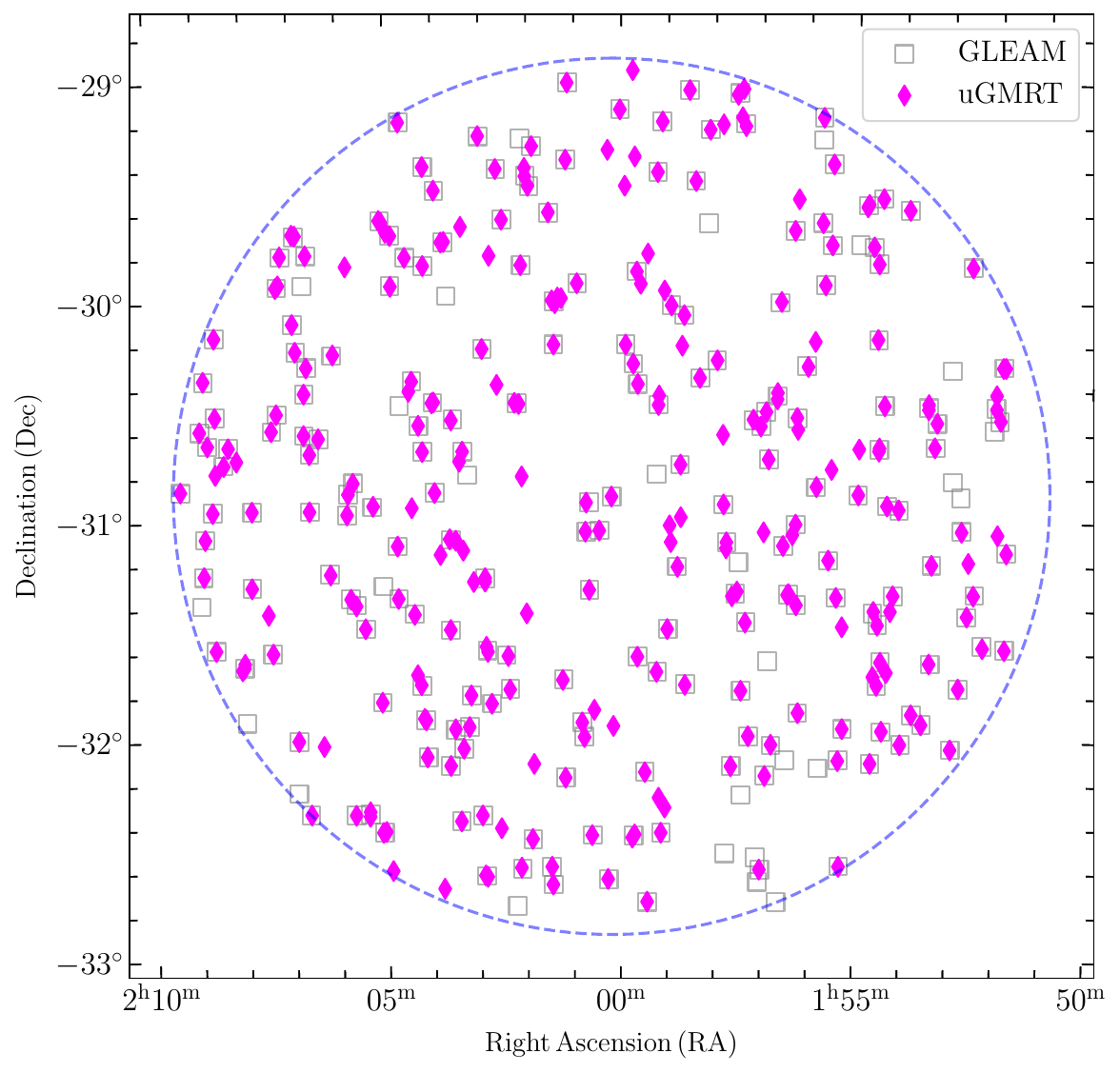}
    \caption{HERA sky model simulations with uGMRT and GLEAM sources. All sources in these two catalogues that are within a $2^\circ$ radius from the center (PC A) are included in the simulations.}
    \label{fig:sky_model_sources_wcs_style}
\end{figure*}

We generate the sky model visibilities for HERA, which are needed to obtain the degeneracy parameters in antenna gain solutions using the absolute calibration. The steps are described in detail in \cite{Kern2020}. At this point, we note that although our sky model is deeper than GLEAM, we do not have the $10^\circ$ sky coverage that covers HERA's FWHM (Section~\ref{sec:ugmrt_observation}). Through this pilot uGMRT survey, we have reached a near-uniform sensitivity of $\approx 2-3$ mJy/beam only in the central $3^\circ$. Therefore, we do not attempt the absolute calibration in this paper. Here, we compare the two catalogues (uGMRT and GLEAM) using only the overlapping part of the sky model. We have already proposed more observation time for a larger survey and plan a detailed absolute comparison in follow-up work.

We choose a $2^\circ$ radius centred on the GLEAM~02H field ($\rm{RA} = 02^{h}\,00^{m}\,12.00^{s}, \rm{Dec} = -30^{\circ}\,53^{\prime}\,23.99^{\prime\prime}$), and include all the sources within this radius in the simulation. We found 296 sources in our catalogue as compared to 239 sources found in GLEAM. Figure~\ref{fig:sky_model_sources_wcs_style} shows all the identified sources in these two catalogues that are within a $2^\circ$ radius from the center. The white squares and pink diamond symbols show the position of the sources detected in the GLEAM and uGMRT catalogues, respectively. We have detected several sources that were missing in GLEAM. We can also identify a few sources that are detected in GLEAM but not by uGMRT. These undetected sources mainly lie either at the periphery of the central region or near bright sources, where our rms sensitivity is poorer. Since we do not have a very accurate measurement of the spectral indices of all the sources, we have assumed all sources to have a constant spectral index of $\alpha=-0.65$, which we found in Section~\ref{sec:spindex}. A follow-up analysis with the GWB data will estimate the in-band spectral indices and make a more detailed comparison.

\begin{table}
    \centering
    \setlength{\tabcolsep}{10pt} % Adjust column spacing
    \renewcommand{\arraystretch}{1.3} % Adjust row spacing
    \caption{Summary of HERA simulation parameters used in this work.}
    \label{tab:hera_sim_params}
    \begin{tabular}{ll}
        \hline
        Parameter & Value \\
        \hline
        Telescope & HERA \\
        Location & Latitude: $-30.7215^\circ$, \\
        & Longitude: $21.4283^\circ$, \\
        & Altitude: 1051.69~m \\
        Number of antennas & 39 \\
        Frequency range & $100-200$~MHz \\
        Number of channels & 1024 \\
        Channel resolution & 97.65625~kHz \\
        Sky model & uGMRT / GLEAM catalogue \\
        Field  & $2^\circ$ centering GLEAM~02H  \\
        LST (mid) & 2 hr \\
        JD (start) & 2458100.295225683 \\
        Integration time & 10.73~s \\
        Observation duration & $5.36$~min (30 timestamps) \\
        Primary beam model & \texttt{PolyBeam} \\
        % (see Table~\ref{tab:beam_yaml}) \\
        \hline
    \end{tabular}
\end{table}

We have used the publicly available  \texttt{hera\_sim}\footnote{\url{https://github.com/HERA-Team/hera_sim/}} codes to simulate the visibilities. The visibilities are simulated using a matrix-based visibility simulator \texttt{matvis} \citep{matvis}. The details of the simulations, which are tabulated in  Table~\ref{tab:hera_sim_params}, are passed to \texttt{matvis} through a \texttt{yaml} file. It includes the uGMRT/GLEAM catalogues, a primary beam model, frequency information, and the antenna layout \citep{HERA2024}. We have used the analytical beam model \texttt{PolyBeam} described in \cite{Choudhuri2021polybeam}.
The simulation is carried out in the frequency range $100-200$~MHz with a channel resolution of 97.66~kHz.  The observation has 30 timestamps, each with an integration time of 10.73~seconds, which leads to a total observation time of $5.36$~min. The Julian Date (JD) is set such that the LST 2~hr (i.e., the GLEAM~02H field) remains overhead at the mid-point of the total observation duration. We have tried to keep the simulation parameters similar to \cite{Kern2020}, and a summary of the parameters is presented in Table~\ref{tab:hera_sim_params}. 

\begin{figure*}
    \includegraphics[width=0.9\textwidth]{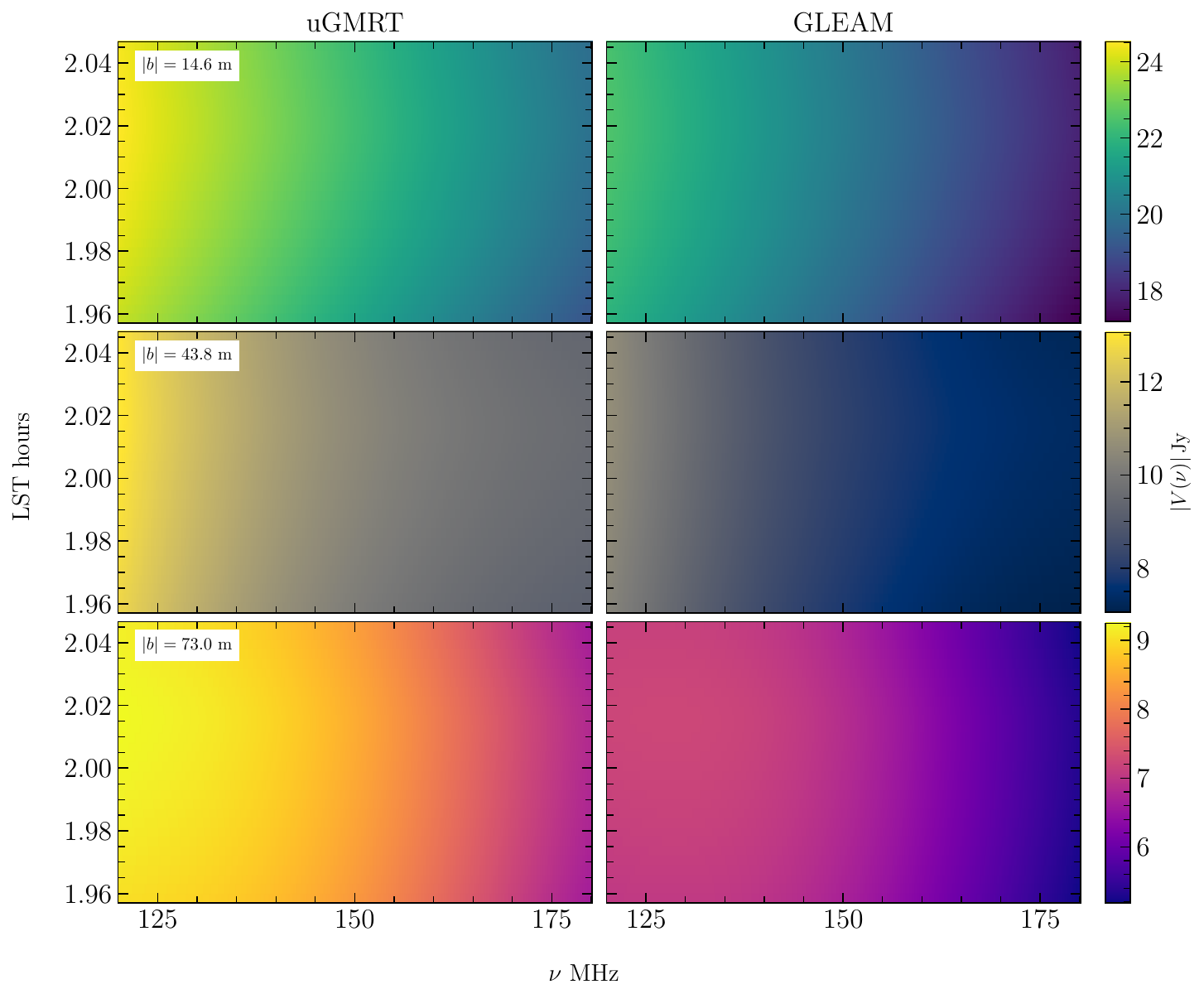}
    \caption{LST-frequency waterfall of the amplitude of the visibilities for three redundant baselines for the sky models with uGMRT and GLEAM catalogues.}
    \label{fig:lst_freq_amp}
\end{figure*}

\begin{figure*}
    \includegraphics[width=0.9\textwidth]{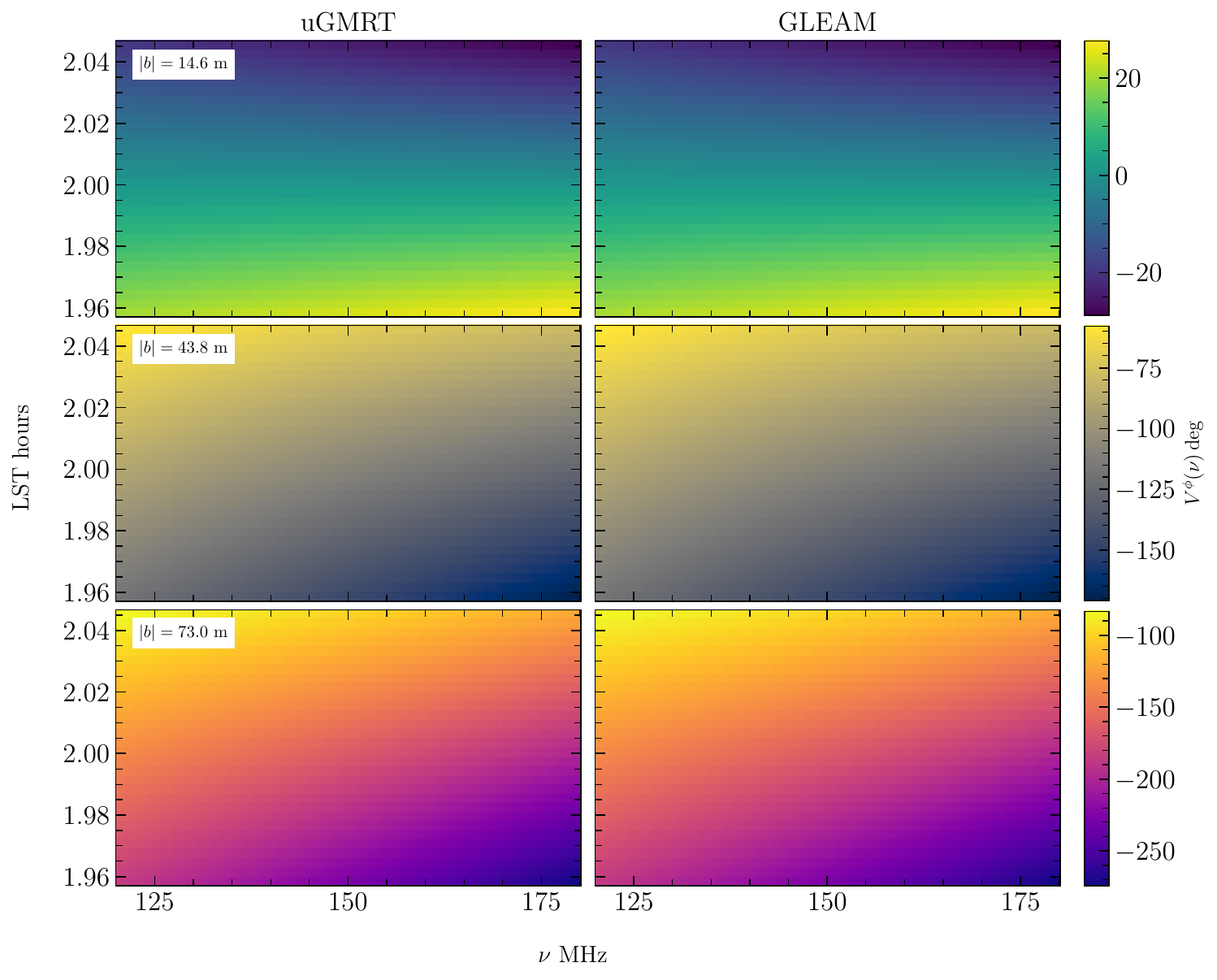}
    \caption{LST-frequency waterfall of the phase of the visibilities for three baselines for the sky models with uGMRT and GLEAM catalogues.}
    \label{fig:lst_freq_phase}
\end{figure*}

Figure~\ref{fig:lst_freq_amp} shows the LST-frequency waterfall of the amplitude of the simulated visibilities for three redundant baselines with $|b| = 14.0\,{\rm m}, 43.8\,{\rm m} \, {\rm and} \, 73.0\,{\rm m}$  for the sky models with uGMRT and GLEAM catalogues. We find that the visibility amplitude gradually decreases over the bandwidth. This is consistent with what we expect for sources with a spectral index of $-0.65$. Further, the sources are directly overhead, and the HERA main lobe does not vary significantly over the frequency range of our simulations.  We do not see much variation in amplitude along LST, which is expected because we have considered a short observation duration ($5.36$~min) and a limited part of the sky (radius$=2^\circ$), and all the sources are expected to remain within the FoV for the entire duration of the observation. Figure~\ref{fig:lst_freq_phase} shows the LST-frequency waterfall of the phase of the visibilities for the same baselines. We see that the phases have not changed much for these two catalogues.

\begin{figure*}
    \includegraphics[width=0.9\textwidth]{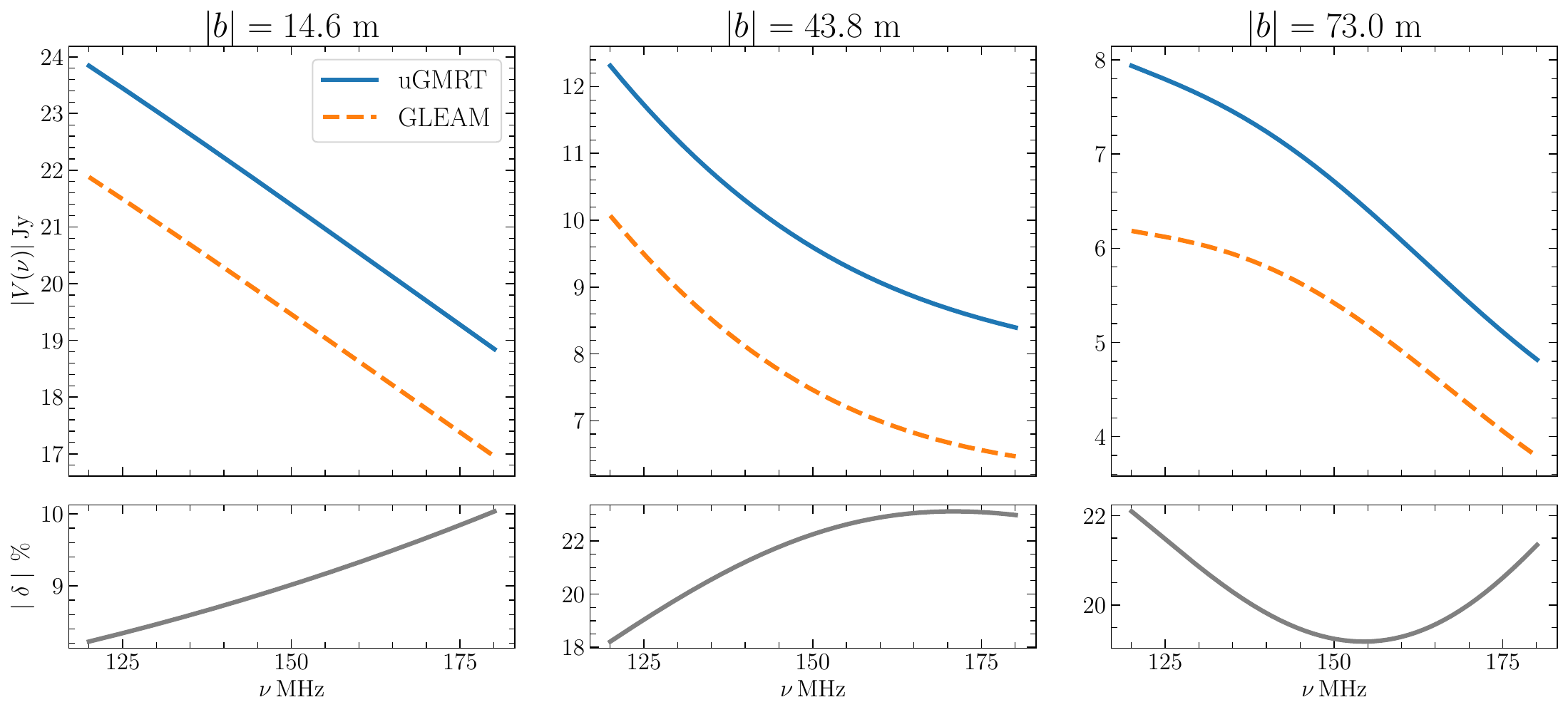}
    \caption{Time averaged visibility amplitudes as a function of frequency for three baselines for the sky models with uGMRT and GLEAM catalogues. The bottom panel shows the percentage difference $\delta = 100\times(|V|_{\rm uGMRT} - |V|_{\rm GLEAM})/|V|_{\rm uGMRT}$.}
    \label{fig:vis_amp_comparison}
\end{figure*}

\begin{figure*}
    \includegraphics[width=0.9\textwidth]{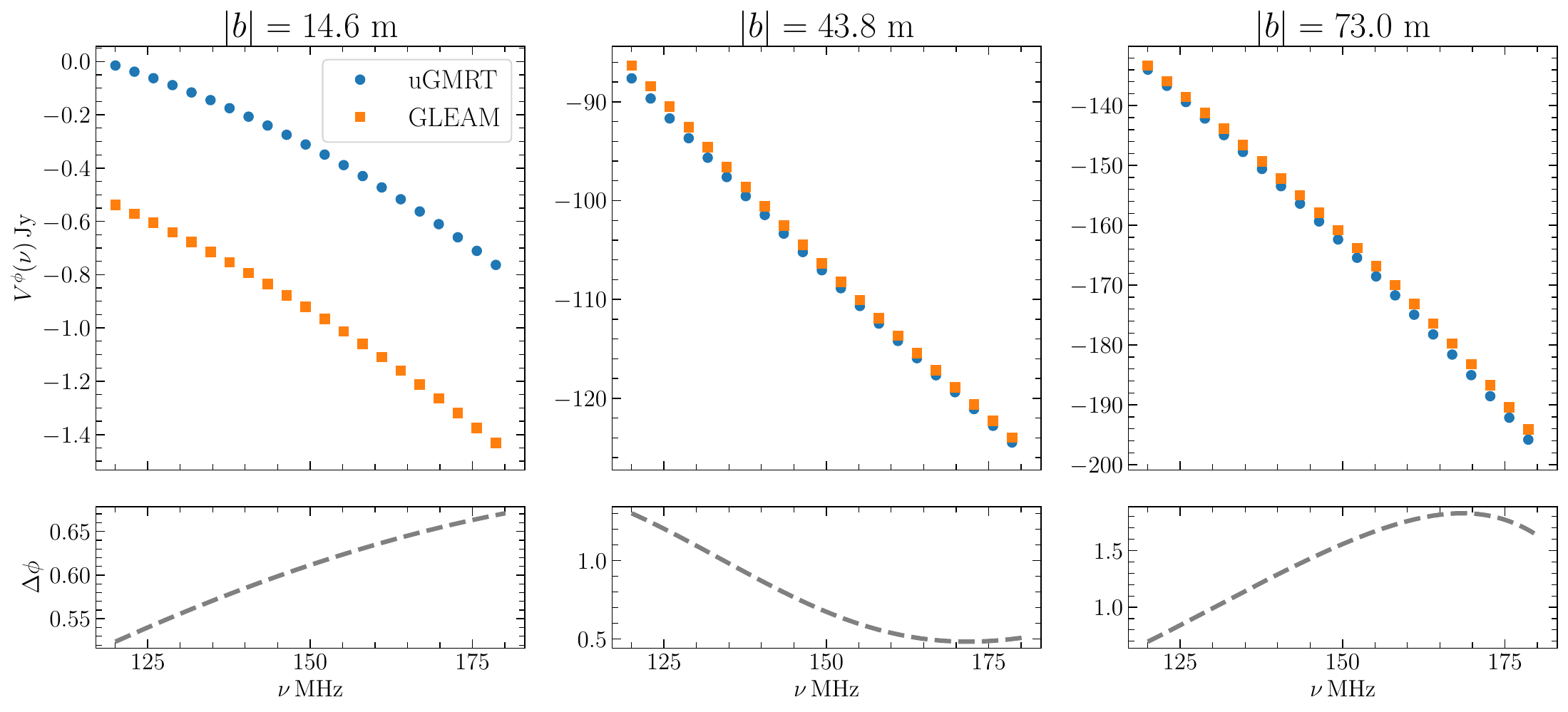}
    \caption{Time averaged visibility phases as a function of frequency for three baselines for the sky models with uGMRT and GLEAM catalogues. The bottom panel shows the difference in the phases $\Delta\phi = |V^\phi_{\rm uGMRT} - V^\phi_{\rm GLEAM}|$.}
    \label{fig:vis_phase_comparison}
\end{figure*}

Since the visibility amplitude does not seem to vary significantly over the observation duration, we average the visibilities in time and compare the time-averaged visibilities. Figure~\ref{fig:vis_amp_comparison} and ~\ref{fig:vis_phase_comparison} respectively show amplitude and phase of the time-averaged visibilities as a function of frequency for those three redundant baselines. We see that the visibility amplitude for uGMRT is higher than that of GLEAM, which is expected as the brightest source in the field is found to be $\approx4\%$ brighter in the uGMRT catalogue (Section~\ref{sec:bright}).
Further, the fluxes for the cross-matched sources are found to be $\approx8\%$ higher in our catalogue (Section~\ref{sec:crossmatch}), which also supports the difference in the amplitude. 
The bottom panel of Figure~\ref{fig:vis_amp_comparison} shows the percentage deviation $\delta = 100\times(|V|_{\rm uGMRT} - |V|_{\rm GLEAM})/|V|_{\rm uGMRT}$ between the visibility amplitude obtained from uGMRT and GLEAM catalogues. Considering the short baselines (e.g. 14.0~m), we find the deviation $\delta \sim 10\%$, which goes up to $25\%$ for a longer baseline (e.g. 73~m).  The bottom panel of Figure~\ref{fig:vis_phase_comparison} shows the difference in the phases of the visibilities $\Delta\phi = |V^\phi_{\rm uGMRT} - V^\phi_{\rm GLEAM}|$ obtained from uGMRT and GLEAM catalogues.  
We find a relatively small phase difference $\Delta\phi \lesssim 2$ considering these baselines.

% \clearpage

\section{Conclusions and Future Plans}
\label{sec:conclusions}

The 21-cm signal from the Epoch of Reionization (EoR) has the potential to reveal the detailed nature of the very first stars and galaxies of our Universe. However, the signal is much fainter than astrophysical foregrounds and requires precise calibration of the interferometric array for its detection. HERA has the sensitivity to measure the EoR 21-cm power spectrum. It uses its specifically designed, nearly identical array layout to perform a redundant calibration, which solves for the antenna gains as a function of time and frequency. However, a reference sky model is needed to fix the degeneracy parameters in the redundant gain solutions. The accuracy and completeness of the sky model are of paramount importance in achieving the near-perfect calibration that is required to detect the EoR 21-cm signal. We have used uGMRT observations to make a high-fidelity source catalogue for the HERA experiment. The observation uses seven pointing centers (PC)  around the GLEAM~02H field to achieve a uniform sky coverage over a $3^\circ$ radius. Here, we present the results from 20 hours of data recorded using GMRT Software Backend (GSB), which has $16.7\,\rm{MHz}$ bandwidth centring $147.4\,\rm{MHz}$. The mosaic of seven PCs (Figure~\ref{fig:mosaic_plot}) has yielded a median rms of $3.9^{+3.7}_{-1.4}$~mJy/beam at $147.4\,\rm{MHz}$, which goes down to $\sim2$~mJy/beam at source-free regions. This is significantly deeper than the GLEAM catalogue, which reported an rms of $10\pm5$~mJy/beam over the region $-72^{\circ} \leq {\rm Dec} < +18.5^{\circ}$ at an even higher frequency of 200~MHz (see Table~4 of \citealt{HW22}).

We use the mosaic to make a catalogue which comprises $640$ sources in the flux range of $0.02-19.08$~Jy. We found $\approx26\%$ of these sources to have extended structures. We cross-matched the sources with TGSS, NVSS, GLEAM, and GLEAM--X catalogues, and found that our uGMRT catalogue has a sub-arcsec positional offset and consistent flux measurements with existing radio catalogues (Figures~\ref{fig:positional_offsets} and \ref{fig:flux_comparison}). We have also computed spectral indices ($\alpha$) for sources cross-matched between our 147.4~MHz uGMRT catalogue and the 1.4~GHz NVSS catalogue, assuming the sources to follow a power-law relation in frequency. The spectral indices are found to span the range $\alpha\approx -1.5$ to $0.5$, with a median value of $\alpha = -0.65^{+0.21}_{-0.21}$ (Figure~\ref{fig:spectral_index}).
We have checked the fluxes and spectral indices of the brightest sources in our catalogue and cross-matched those values with TGSS, NVSS, GLEAM, and GLEAM--X (Table~\ref{tab:gmrt_gleam_nvss_alpha}). 
We find that both the flux and spectral index measurements are somewhat close. However, accurate measurements of the spectral index of the fainter sources would need further investigations, e.g., using the GMRT Wideband Backend (GWB) data ($120-250$~MHz) and other uGMRT bands such as Band 3 ($250-500$~MHz). 

We have used the catalogue to measure the differential source counts and find it to be deeper than GLEAM source counts and consistent with LoTSS-based counts at $\sim150$~MHz (Figure~\ref{fig:source_counts_comparison}). This demonstrates that deeper catalogues can be made with complementary arrays such as uGMRT to have a complete sky coverage of the HERA calibration fields. Although our sky model is deeper than GLEAM, the present pilot survey does not have the $10^\circ$ sky coverage that covers HERA's FWHM. Here, we have reached a near-uniform sensitivity of $\approx 2-3$ mJy/beam only in the central $3^\circ$. Based on the results from this pilot study, we have proposed (ID: $49\_019$) more observation time for a larger survey, and another 40~hours of observation time has been awarded for this purpose. 

We have carried out preliminary simulations of the sky models for HERA absolute calibration using our uGMRT and existing GLEAM catalogues, in the central $2^\circ$ region of the GLEAM~02H, where our sensitivity is nearly uniform. The preliminary analysis assumes a constant spectral index of $\alpha = -0.65$ for all the sources. We see a $10-25\%$ difference in the visibility amplitudes between these two catalogues, with a relatively small change ($\approx2^\circ$) in phases. We plan to analyze the full GWB data ($120-250$~MHz) to reduce the rms and measure the in-band source spectral indices of fainter sources, which is expected to improve the fidelity of the HERA calibration model. We plan a detailed absolute comparison in follow-up work. We also plan to employ multiple uGMRT bands for precise measurement of spectral indices in the future.

\section*{Acknowledgements}

We appreciate the careful reading and constructive comments of the Scientific Editor and the anonymous Reviewer, which have helped us improve the manuscript. We thank the staff of the GMRT, which is operated by the National Centre for Radio Astrophysics of the Tata Institute of Fundamental Research, for making these observations possible. AE and SC acknowledge the support from the CoE research grant, IIT Madras, through funding and computational resources. AE thanks Dr Somnath Bharadwaj and Dr Shiv Sethi for their support through visits, and also thanks Dr Suman Chatterjee, Dr Arnab Chakroborty, and Rashmi Sagar for their useful comments. AE also acknowledges the use of \textsc{ChatGPT} for assistance with plotting and debugging. SC would also like to acknowledge the SERB Start-up Research Grant (File No: SRG/2023/000622) for providing financial support. We acknowledge the International Centre for Theoretical Sciences (ICTS), which facilitated valuable discussions through the program, Radio Cosmology and Continuum Observations in the SKA Era: A Synergic View (code:  ICTS/radiocoscon2025/04). This result is part of a project that has received funding from the European Research Council (ERC) under the European Union's Horizon 2020 research and innovation programme (Grant agreement No. 948764; PB). This research has made use of the \textit{VizieR}\footnote{\url{https://vizier.cds.unistra.fr/}} catalogue access tool, CDS, Strasbourg, France \citep{VizieR2000}.
% The source catalogues used in this paper have been obtained through \textit{VizieR}\footnote{\url{https://vizier.cds.unistra.fr/}}\citep{VizieR2000}}.
The following softwares were used in this work: \textsc{Astropy} \citep{astropy13, astropy18}, \textsc{SPAM} \citep{Intema2009}, \textsc{AIPS} \citep{Greisen1998, Greisen2002}, \textsc{PyBDSF} \citep{PyBDSF},  \textsc{aegean} \citep{Hancock2012aegean1, Hancock2018aegean2}, and several \textsc{Python} packages, including \textsc{Numpy}, \textsc{Scipy}, and \textsc{Matplotlib}.

%%%%%%%%%%%%%%%%%%%%%%%%%%%%%%%%%%%%%%%%%%%%%%%%%%
\section*{Data Availability}

The observed data are publicly available through the GMRT data archive\footnote{\url{https://naps.ncra.tifr.res.in/goa/}} under the proposal code $43\_119$. The calibrated data and the simulated data used here are available upon reasonable request to the corresponding author. The codes used for the analysis are available on GitHub\footnote{\url{https://github.com/aecosmo/RADIOcat}}. 
The catalogue can be found as a FITS table in the supplementary material and will be available at CDS via anonymous ftp to cdsarc.u-strasbg.fr (130.79.128.5) or via \url{https://cdsarc.cds.unistra.fr/viz-bin/cat/J/MNRAS}.

% The inclusion of a Data Availability Statement is a requirement for articles published in MNRAS. Data Availability Statements provide a standardised format for readers to understand the availability of data underlying the research results described in the article. The statement may refer to original data generated in the course of the study or to third-party data analysed in the article. The statement should describe and provide means of access, where possible, by linking to the data or providing the required accession numbers for the relevant databases or DOIs.



\bibliographystyle{mnras}
\bibliography{example} % if your bibtex file is called example.bib

%%%%%%%%%%%%%%%%%%%%%%%%%%%%%%%%%%%%%%%%%%%%%%%%%%

%%%%%%%%%%%%%%%%% APPENDICES %%%%%%%%%%%%%%%%%%%%%

\appendix

\section{uv-coverage and PSF}
\label{sec:uvtracks}

\begin{figure*}
    \centering
    \includegraphics[width=\textwidth]{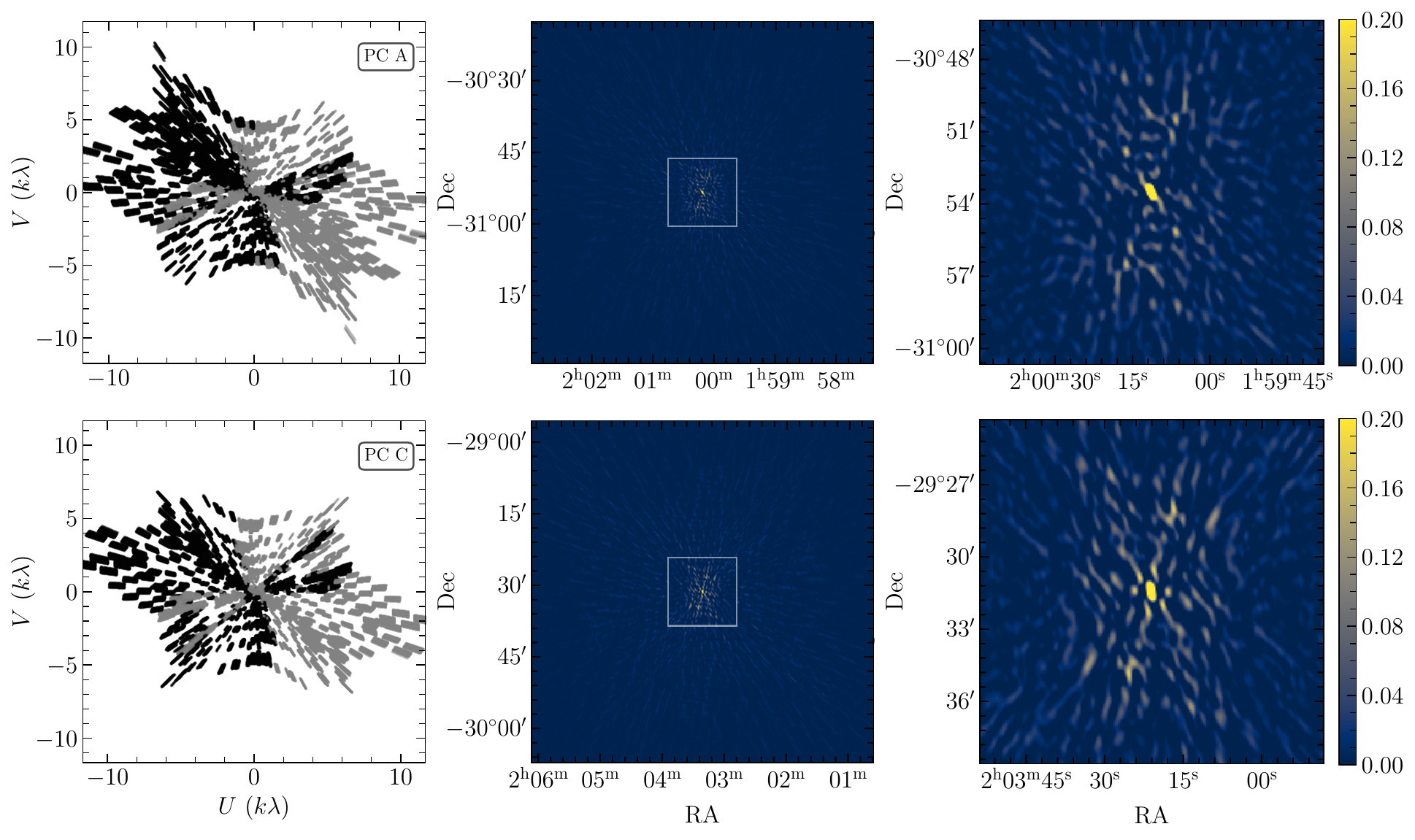}
    \caption{The $uv$-coverage (left panels), the corresponding PSF (middle panels), and the central one-fifth part of the PSF (right panels) for the PCs A and C. To emphasize the sidelobe structure of the PSF, the color scale is saturated at 0.20, which is approximately the maximum sidelobe level. Other PCs exhibit similar $uv$-coverage and PSF patterns.}
    \label{fig:uvtracksandpsf}
\end{figure*}

The extent and uniformity of the $uv$-coverage, and thereby the point spread function (PSF), directly influence the image fidelity, noise characteristics, and dynamic range achievable in the images. Figure~\ref{fig:uvtracksandpsf} shows the $uv$-coverage and the corresponding PSF for the PCs~A and C. The middle panels show the full PSF, which is further zoomed into the rightmost panel to highlight its sidelobe structures. Other PCs also exhibit similar $uv$-coverage and PSF patterns. 

\section{Source finding and background rms estimation}
\label{sec:pybdsf}

We have made use of the software Python Blob Detection and Source Finder (\textsc{PyBDSF}; \citealt{PyBDSF}) at different stages of the image analysis to perform three main tasks -- estimating background rms noise map of an image, finding sources, and making a catalogue. Here, we describe the `tasks' and parameter choices we made in \textsc{PyBDSF}, which are used throughout the work. 

\begin{table}
    \centering
    \caption{\textsc{PyBDSF} parameter settings used for source extraction and background rms estimation.}
    \label{tab:pybdsf_params}
    \begin{tabular}{l c p{4.cm}}  % Adjust width as needed
        \hline
        Parameter & Value & Description \\
        \hline
        \texttt{adaptive\_rms\_box} & True & Enables adaptive rms box-sizing to handle spatial noise variations. \\
        \texttt{rms\_box} & (150, 30) & Large box size with a small step to ensure robust noise estimation. \\
        \texttt{rms\_box\_bright} & (50, 10) & Smaller box for bright regions to improve local noise estimation. \\
        \texttt{psf\_vary\_do} & True & Allows for PSF variation across the image for better Gaussian fitting. \\
        \texttt{thresh\_isl} & 3.0 & Island threshold at 3$\sigma$ to ensure reliable source detection. \\
        \texttt{thresh\_pix} & 5.0 & Pixel threshold at 5$\sigma$ to minimize false detections. \\
        \hline
    \end{tabular}
\end{table}

The main task in \textsc{PyBDSF} is \texttt{process\_image}, which reads an input image (e.g., those obtained from SPAM), calculates background rms and mean maps, finds islands of emission, fits Gaussians to the islands, and groups the Gaussians into sources. The software allows one to choose parameters to carry out these actions, and the details can be found in the software's documentation\footnote{\url{https://pybdsf.readthedocs.io/}}. Table~\ref{tab:pybdsf_params} provides a brief description of our parameter choice, which we have used uniformly throughout the paper. 

A key parameter for estimating the background rms of an image is  \texttt{rms\_box}, which has two inputs. The first is boxsize, which is the size of the two-dimensional (2D) sliding box for calculating the rms and mean over the entire image. The second, stepsize, is the number of pixels by which this box is moved for the next measurement. We have set the \texttt{rms\_box} to be adaptive (\texttt{adaptive\_rms\_box = True}) to handle spatial noise variations in the images. This choice ensures that the \texttt{rms\_box} is reduced in size near bright sources and enlarged far from them. This scaling attempts to account for possible strong artifacts around bright sources while still achieving accurate background rms and mean values when extended sources are present. For generating the rms map, we have set the large-scale (box size, step size) to be \texttt{rms\_box = (150, 30)}, while near bright sources, it is set as \texttt{rms\_box\_bright = (50, 10)}. We have used \texttt{psf\_vary\_do = True} to capture the variation of the PSF across an image for better Gaussian fitting. We have kept all the sub-parameters under it at the default. For identifying the bright sources in the image, we have used a threshold of $3\sigma$ to detect islands (\texttt{thresh\_isl = 3}) and $5\sigma$ for source detection (\texttt{thresh\_pix = 5}). Throughout the paper, unless explicitly mentioned, we use \textsc{PyBDSF} to generate the rms map of an image, while excluding bright outlier sources with flux densities greater than five times the local rms. The median value of this rms map is then quoted as the representative rms of the image.

\section{Definition of the columns}
\label{sec:catalog_tab_def}

Table~\ref{tab:catalogue_columns} provides the definitions of all 45 columns of the uGMRT 147.4~MHz source catalogue that is provided as a FITS file in the supplementary material. This catalogue represents the source list (`srl') output of the \texttt{write\_catalog} task of \textsc{PyBDSF}. A detailed description of all the columns can be found in \textsc{PyBDSF}'s documentation\footnote{\url{https://pybdsf.readthedocs.io/en/latest/write_catalog.html}}. 

\begin{table*}
    \centering
    \caption{Definitions of all 45 columns in the uGMRT 147~MHz Source Catalogue (FITS). The Column numbers, labels, units, and a brief explanation are provided. Column numbers, labels, units, and brief explanations are provided. Columns labeled with errors correspond to $1\sigma$ uncertainties. All positions are given in the J2000 equinox. The position angles (PA) are measured east of north.}
    \label{tab:catalogue_columns}
    \begin{tabular}{clll}
    \hline
    Column & Label & Unit & Explanation \\
    \hline
    1  & Source\_id         & ---         & Unique ID for the source \\
    2  & Isl\_id            & ---         & Unique ID for the island \\
    3  & RA                 & deg         & Right ascension of the source \\
    4  & E\_RA              & deg         & Error on RA \\
    5  & DEC                & deg         & Declination of the source \\
    6  & E\_DEC             & deg         & Error on DEC \\
    7  & Total\_flux        & Jy          & Total integrated flux of the source \\
    8  & E\_Total\_flux     & Jy          & Error on total flux \\
    9  & Peak\_flux         & Jy/beam     & Peak Stokes I flux density per beam \\
    10 & E\_Peak\_flux      & Jy/beam     & Error on peak flux density \\
    11 & RA\_max            & deg         & RA of the maximum of the source \\ 
    12 & E\_RA\_max         & deg         & Error on RA of the maximum of the source \\
    13 & DEC\_max           & deg         & DEC of the maximum of the source \\
    14 & E\_DEC\_max        & deg         & Error on DEC of the maximum of the source \\
    15 & Maj                & deg         & FWHM of the major axis of the source \\
    16 & E\_Maj             & deg         & Error on major axis \\
    17 & Min                & deg         & FWHM of the minor axis \\
    18 & E\_Min             & deg         & Error on minor axis \\
    19 & PA                 & deg         & PA of major axis \\
    20 & E\_PA              & deg         & Error on position angle \\
    21 & Maj\_img\_plane    & deg         & Major axis in image plane \\
    22 & E\_Maj\_img\_plane & deg         & Error on major axis in image plane \\
    23 & Min\_img\_plane    & deg         & Minor axis in image plane \\
    24 & E\_Min\_img\_plane & deg         & Error on minor axis in image plane \\
    25 & PA\_img\_plane     & deg         & PA in image plane \\
    26 & E\_PA\_img\_plane  & deg         & Error on PA in image plane \\
    27 & DC\_Maj            & deg         & FWHM of the deconvolved major axis of the source \\
    28 & E\_DC\_Maj         & deg         & Error on deconvolved major axis \\
    29 & DC\_Min            & deg         & FWHM of the deconvolved minor axis of the source \\
    30 & E\_DC\_Min         & deg         & Error on deconvolved minor axis \\
    31 & DC\_PA             & deg         & PA of the deconvolved major axis of the source \\
    32 & E\_DC\_PA          & deg         & Error on deconvolved position angle \\
    33 & DC\_Maj\_img\_plane & deg        & Deconvolved major axis in image plane \\
    34 & E\_DC\_Maj\_img\_plane & deg     & Error on deconvolved major axis in image plane \\
    35 & DC\_Min\_img\_plane & deg        & Deconvolved minor axis in image plane \\
    36 & E\_DC\_Min\_img\_plane & deg     & Error on deconvolved minor axis in image plane \\
    37 & DC\_PA\_img\_plane & deg         & Deconvolved PA in image plane \\
    38 & E\_DC\_PA\_img\_plane & deg      & Error on deconvolved PA in image plane \\
    39 & Isl\_Total\_flux   & Jy          & Total flux of island \\
    40 & E\_Isl\_Total\_flux & Jy         & Error on total flux of island \\
    41 & Isl\_rms           & Jy/beam     & Average background RMS in island \\
    42 & Isl\_mean          & Jy/beam     & Average background mean in island \\
    43 & Resid\_Isl\_rms    & Jy/beam     & Residual RMS in island after Gaussian subtraction \\
    44 & Resid\_Isl\_mean   & Jy/beam     & Residual mean in island after Gaussian subtraction \\
    45 & S\_Code            & ---         & Source structure code: `S'=single, `C'=single in island, `M'=multi-Gaussian \\
    \hline
    \end{tabular}
\end{table*}

%%%%%%%%%%%%%%%%%%%%%%%%%%%%%%%%%%%%%%%%%%%%%%%%%%
\balance

% Don't change these lines
\bsp	% typesetting comment
\label{lastpage}
\end{document}